\documentclass[twocolumn,nofootinbib,showpacs,prd,superscriptaddress]{revtex4}

\usepackage{graphics,graphicx,rotating,color}
\usepackage[percent]{overpic}
\usepackage{mathrsfs,amsmath,amssymb}

\newcommand{\beq}{\begin{equation}}
\newcommand{\eeq}{\end{equation}}
\newcommand{\bea}{\begin{eqnarray}}
\newcommand{\eea}{\end{eqnarray}}

\newcommand{\lie}{\mathcal{L}}

\begin{document}

\title{	
Wormholes and trumpets: \\
the Schwarzschild spacetime for the moving-puncture generation
}

\author{Mark Hannam} 
\affiliation{Theoretical Physics Institute, University of
  Jena, 07743 Jena, Germany} 
  \affiliation{Physics Department, University
  College Cork, Cork, Ireland} 
\author{Sascha Husa} 
\affiliation{Max-Planck-Institut f\"ur
  Gravitationsphysik, Albert-Einstein-Institut, Am M\"uhlenberg 1, 14476 Golm,
  Germany} 
\author{Frank Ohme}
 \affiliation{Theoretical Physics Institute, University of
  Jena, 07743 Jena, Germany} 
\author{Bernd Br\"ugmann} 
\affiliation{Theoretical Physics Institute,
  University of Jena, 07743 Jena, Germany} 
\author{Niall \'O~Murchadha}
\affiliation{Physics Department, University
  College Cork, Cork, Ireland}

\date{\today}

\begin{abstract}
We expand upon our previous analysis of numerical moving-puncture simulations 
of the Schwarzschild spacetime. We present a derivation of the family
of analytic stationary 1+log foliations of the Schwarzschild solution,
and outline a transformation to isotropic-like coordinates.
We discuss in detail the numerical evolution of
standard Schwarzschild puncture data, and the new time-independent 1+log
data. Finally, we demonstrate that the moving-puncture method can locate the
appropriate stationary geometry in a robust manner when a numerical code 
alternates between two forms of 1+log slicing during a simulation.
\end{abstract}

\pacs{
04.20.Ex,   % Initial value problem, existence and uniqueness of solutions
04.25.Dm, % Numerical relativity
04.30.Db, % Wave generation and sources (Gravitational wave theory)
95.30.Sf    % Relativity and gravitation (Fundamental aspects of astrophysics)
}

\maketitle

\section{Introduction}

The binary black hole problem is a cornerstone problem in gravitational
theory: solving for the inspiral and merger of two black holes in full general
relativity connects the theory's strong-field regime with astrophysics,
and connects issues with the mathematical
understanding of the theory with the emerging field of gravitational-wave
astronomy. Solutions of the binary black hole problem require numerical
simulations, but stable, long-term numerical evolutions of orbiting black-hole 
binaries eluded researchers for four decades. 
However, after a number of insights and technical developments
two independent methods \cite{Pretorius:2005gq,Campanelli:2005dd,Baker05a}
were shown in 2005 to allow simulations of the last orbits, merger and ringdown of
an equal-mass nonspinning binary. One of these,
the ``moving puncture method'' {\cite{Campanelli:2005dd,Baker05a}},  was quickly
adopted by many research groups and has since been applied to many, more
general, scenarios: unequal-mass binaries
{\cite{Herrmann2006,Baker:2006nr,Gonzalez06tr,Buonanno:2007pf}}, 
spinning binaries
{\cite{Campanelli2006b,Campanelli2006c,Campanelli2007,Campanelli:2006fy,Herrmann2007,Herrmann:2007ex,Gonzalez:2007hi,Hannam:2006zt,Brugmann:2007zj,Koppitz-etal-2007aa,Pollney:2007ss,Rezzolla:2007xa,Rezzolla:2007rd,Hannam:2007wf,Dain:2008ck,Baker:2008md}},
three-black-hole spacetimes \cite{Campanelli:2007ea,Lousto:2007rj}
and long simulations of (so far) up to ten orbits
{\cite{Baker2006a,Hannam:2007ik,Husa2007a,Husa:2007ec}}.

At the technical level, the moving-puncture method consists of a
seemingly small modification of the earlier ``fixed puncture''
method~\cite{Bruegmann97,Alcubierre02a}. However, 
in~\cite{Hannam:2006vv} (which we will refer to as Paper I), we found
that the numerical slices behave in a way radically different from
what was observed in previous fixed-puncture evolutions,
and indeed in all previous numerical simulations of black-hole spacetimes. 
We developed a {\it geometrical} picture of the behavior of moving-puncture
simulations, and our results suggested a new paradigm for black-hole evolutions,
centered around manifestly stationary representations of black-hole spacetimes
and ``asymptotically cylindrical'' slices. This led to further investigations 
in {\cite{Hannam:2006xw,Brown:2007tb,Garfinkle:2007yt}}.  
In this paper we expand and discuss in more depth the results of Paper I, with
particular reference to the asymptotics of the initial and final
states of a moving-puncture simulation

The initial data in a typical moving-puncture simulation represent black holes
using a wormhole topology: as we follow the coordinates toward one
of the black holes, we do not reach the black hole's singularity but instead
pass through a wormhole to another exterior space, and eventually find
ourselves once again in an asymptotically flat region. Data for $N$ black
holes can consist of $N + 1$ asymptotically flat regions connected by $N$
wormholes. In the puncture approach to constructing initial data
{\cite{Beig94,Beig:1994rp,Brandt97b,Dain01a}} each ``extra''
asymptotically flat region is compactified so that its spatial
infinity is transformed to a single point (``puncture''), and the
entire $N + 1$-wormhole topology can be represented in a single
three-dimensional space, $\mathbb R^3$. All of the black-hole singularities
are conveniently avoided in this construction, and there is no need to
``excise'' any region when these data are used in a numerical
simulation.

This use of nontrivial topology to enforce the presence of horizons can be
regarded as a mere {\em trick} to conveniently construct black-hole
initial data. It comes at the expense of using the entire Kruskal extension to
the Schwarzschild spacetime, part of which has no physical relevance in
typical numerical evolutions. Our analysis suggests that a similar trick could
be used to construct initial data that leave out most of the unphysical region;
we will later refer to these as ``trumpet'' data.

Standard puncture data are smooth over the entire space, except for a
scalar function, the conformal factor $\psi$, which diverges as $1/r$
near each puncture. 
One of the two innovations of the moving-puncture
method was that it provides a method to stably evolve the conformal factor. In
addition, the method uses gauge conditions (variants of ``1+log'' slicing for
the lapse function {\cite{Bona1995}} and $\tilde{\Gamma}$-freezing for the
shift vector \cite{Alcubierre01a,Alcubierre02a}) that allow the punctures to
move across the numerical grid. The result is that the punctures orbit each
other and spiral inwards, as if the black holes were being represented by
point particles, and the plots of the ``puncture tracks'' shown in many papers
easily match our intuitive picture of objects in orbit (see, for example,
{\cite{Campanelli:2005dd,Baker05a,Baker2006a,Husa2007a}}).
Of course, the orbiting punctures are {\em not} point particles, they are the
asymptotic infinities of wormholes, or at least they were in the initial data.
Are these two extra copies of an asymptotically flat
region of spacetime orbiting each other on the numerical grid?
The answer turns out to be No, as we explain below.

In Paper I we studied moving-puncture simulations of a Schwarzschild
black hole, and found that the numerical points quickly leave the
extra copy of the exterior space. 
Where the points near the puncture originally approached a second
asymptotically flat end, after a short time the grid points instead
asymptote to an infinitely long ``cylinder'' with radius $R_0 = 1.31M$.
In addition, we solved the spherically symmetric Einstein equations
for a stationary 1+log-sliced spacetime (see also {\cite{Buchman:2005ub}}), 
and found that the
one solution with a lapse function that is non-negative everywhere
agrees with that found by the numerical code. In other words, there is
one regular stationary 1+log solution, and the moving-puncture method
quickly finds it. Whereas an embedding diagram of the initial data
resembles a wormhole, we will refer to that of the late-time slices as
a ``trumpet''. This terminology becomes clear in
Figures~\ref{fig:emb_wormhole} and \ref{fig:emb_trumpet}.

This seemingly dramatic change in the appearance of the numerical
slices is achieved by the gauge conditions.  We start with
moment-of-time-symmetry initial data for the Schwarzschild solution,
choose the initial lapse $\alpha \equiv 1$ and propagate using the
1+log  condition. So far the configuration is symmetric across the
throat; we will call this ``left-right'' symmetry to be consistent
with standard Kruskal and Penrose diagrams, although we mean symmetry
with respect to the upper and lower halves of
Figure~\ref{fig:emb_wormhole}. So: the initial configuration is
left-right symmetric, and the slicing condition preserves this
symmetry. This means that all the slices of the foliation will also be
left-right symmetric and run from one spacelike infinity to the
other. As such, there will not be a stationary limiting slice; 
we will elaborate on this point in Section~\ref{sec:analytic}.    
However, at late times a {\it region} of the slice to the right (and a
corresponding region to the left) is asymptotically stationary.  Now
we have to consider the effect of the $\tilde{\Gamma}$-driver shift
condition. If we use puncture data, this condition generates a very
large shift, pointing to the right, near the puncture. This has the
effect of dragging all the data points near the puncture into the
region of the Schwarzschild solution between the horizons and finally
onto the stationary part of the slicing.  This happens extremely
quickly. The closer the innermost data point is to the puncture, the
longer it takes, but it always happens. 
A description of this phenomenon was also given in \cite{Brown:2007tb}. 

There exists a true stationary 1+log slice through the Schwarzschild
spacetime. This is what we call the ``trumpet''. This is
asymptotically flat at one end and cylindrical (of radius $R \approx
1.31 M$) at the other. The asymptotically stationary part of the
wormhole foliation asymptotes to (part of) this trumpet.  The closer
the data points are to the puncture, the more of the trumpet we will
finally see. Further, we will see only  the trumpet --- there are no
longer any grid points on the non-stationary part of the slice, or any 
of the slice on the left. We may say that the left half of the slice
is grossly under-resolved (in fact, it is not resolved at all!), but
the nature of the asymptotic 1+log slice is such that a new
asymptotics has formed, and the left half of the slice is causally
disconnected from the right: the non-resolution of part of the slice
no longer concerns us. 

An example that closely mimics some of the behaviour described above
was found in one of the first ever numerical simulations of a
black-hole spacetime, as early as 1973, by Estabrook {\it et
  al.}~\cite{Estabrook73}. They evolved the Schwarzschild solution
with maximal slicing.  The slices were reflection symmetric about a 
``throat''. At late times the left and right halves looked
approximately stationary. In fact the right half was translated to the
right by the Killing vector while the left half was dragged to the
left. The throat region approximated a cylinder of radius $3M/2$,
which grew linearly with time.  

Two phenomena associated with their simulations --- ``collapse of the
lapse'' and ``slice stretching'' --- were recurring topics in
numerical studies of black-hole spacetimes for the next thirty years 
\cite{Smarr1978,Shapiro86,Beig98,Reimann:2004yf}. 
However, what we consider in this context to be the key feature of
their result, the formation of a cylindrical asymptotics, 
received little, if any, attention before the work in Paper I. 
There we suggested that the existence of a
time-independent representation of the foliation consistent with the
new asymptotics, which a numerical code can find with the aid of an
appropriate shift condition, is one of the keys to the success of the
moving-puncture method. We also took the additional intuitive
step of realizing that this time-independent representation could be
utilized from the outset of a simulation, to allow the construction of 
fully time-independent trumpet puncture data. 

In this paper we extend the 
analysis that we provided in Paper I. In Section~\ref{sec:analytic} we
discuss standard wormhole puncture data, which cannot be
time-independent, and trumpet puncture data, which can.  The canonical
examples are the maximally sliced solution  
{\cite{Estabrook73,Reinhardt73}} and our stationary 1+log solution. Here we
provide a simple derivation of the stationary 1+log solution using a
height function approach, and illustrate some of its features in a
Penrose diagram. We also transform this solution to isotropic
coordinates, providing puncture trumpet data that should be time
independent in a moving-puncture simulation.

In Section~\ref{sec:numerics} we present our numerical method. We start with a
brief description of the main features of the BSSN/moving-puncture system, and
outline a procedure to construct Penrose diagrams from numerical simulations
of the Schwarzschild spacetime. Section~\ref{sec:numresults} contains the
first set of numerical results: a detailed study of a moving-puncture
evolution of wormhole puncture initial data. We show how quickly the numerical
data approach a trumpet geometry, how they behave during the transition, and
check the accuracy with which they approach the analytic stationary solution.
It is important to emphasize that although the numerical solution is
stationary in the sense of coordinate-independent functions (for example, a
plot of $\alpha$ vs $R$ or of $\operatorname{Tr} (K)$ vs $\alpha$),
the numerical coordinates may still exhibit some drift.

This is emphasized in Section~\ref{sec:trumpetevolution}, where we evolve the
{\em trumpet} puncture data produced in Section~\ref{sec:analytic}. We
demonstrate that these are indeed time independent, up to small numerical
errors. Furthermore, if we alternate between two variants of 1+log slicing
during an evolution, we show that the numerical data can alternate between two
stationary solutions, but that some coordinate drift develops in the process.
This drift is minimized by choosing $\eta = 0$ in the
$\tilde{\Gamma}$-driver shift condition.

Finally, we conclude with some remarks about the potential usefulness of
trumpet puncture initial data for black-hole binaries, which will be the
subject of future work.

\section{Wormholes and trumpets: analytic treatment}
\label{sec:analytic}

We begin by summarizing our notation and approach, which is based on
the standard ``3+1'' space+time splitting  of Einstein's equations
\cite{York79}. We introduce the Schwarzschild metric in Schwarzschild
coordinates, progress to isotropic coordinates, which are better
adapted to the standard ``puncture'' method, and then on to a
derivation of the ``trumpet'' solution previously introduced in Paper
I. Our focus here is on solutions that are time-independent, because
even in the black-hole binary problem we want to choose coordinates
that minimize the gauge dynamics; the only dynamics we really want to
see in a numerical code are {\it physical} dynamics.

 \subsection{3+1 decomposition and variables}

We start by making a space plus time (3+1) decomposition of the spacetime
metric, 
\beq
ds^2 = - \alpha^2 dt^2 + \gamma_{ij} (dx^i + \beta^i dt) (dx^j + \beta^j dt).
\eeq
The three-dimensional metric of a $t = {\rm const}$ slice is denoted by
$\gamma_{ij}$. The lapse function $\alpha$ gives the proper time between the
slice at time $t$, and the next slice at time $t + dt$. The shift vector
$\beta^i$ prescribes how the coordinates shift between the two slices. The
complete data on one timeslice are given by $\gamma_{ij}$ and the extrinsic
curvature \beq
K_{ij} = \frac{1}{2\alpha} ( \nabla_i \beta_j + \nabla_j \beta_i - \partial_t
\gamma_{ij} ),
\eeq where $\nabla_i$ denotes covariant differentiation with respect to the
spatial metric $\gamma_{ij}$, and we have used the MTW/ADM sign convention
\cite{MTW,Arnowitt1962}, 
not that given in Wald \cite{Wald84}. It proves convenient to also 
split the extrinsic
curvature into its trace $K$ and tracefree part $A_{ij}$, i.e., \beq
K_{ij} = A_{ij} + \frac{1}{3} \gamma_{ij} K.
\eeq

We further decompose these data with respect to a conformal
metric, $\tilde{\gamma}_{ij}$, as follows: \begin{eqnarray}
\gamma_{ij} & = & \psi^4 \tilde{\gamma}_{ij}, \\
A_{ij} & = & \psi^{-p} \tilde{A}_{ij}, \label{eqn:Adecomp} \\
K & = &  \tilde K, \\
\beta^i & = & {\tilde\beta}^i.
\end{eqnarray} Note that in the standard conformal decomposition often
used to solve the constraint equations \cite{York79,Choquet80}, one chooses $p
= -2$, but for the BSSN decomposition used in the
moving-puncture method, $p = 4$. Note also that the trace of the
extrinsic curvature and the contravariant components of the shift
vector are unchanged by the conformal rescaling. The lapse function
can also be transformed, but this is not necessary for the present
study. We could also have made other choices of the conformal weights,
but we are not interested in those here. 
The complete data on one time slice are $(\gamma_{ij},K_{ij})$, or
equivalently $(\psi,\tilde{\gamma}_{ij},\tilde{A}_{ij},K)$. 

For the moment we will not worry about the details of the constraint
or evolution equations. In this section we will be interested in
finding time-independent representations of the Schwarzschild
solution, for which the constraint equations are already satisfied,
and the time evolution of the data is trivial (i.e., they do not
change in time). Although we will make comments about numerical
simulations in this section, we will postpone concrete details until
Section~\ref{sec:numerics}, when we study numerically the time
development of our data.

\subsection{Wormhole puncture data for a Schwarzschild black hole}

The Schwarzschild metric in Schwarzschild coordinates is 
\beq
ds^2 = - f dT^2 + f^{-1} dR^2 + R^2 d\Omega^2,  \label{eqn:schwarzschild}
\eeq where $f = 1 - 2M/R$. Throughout this paper $R$ and $T$ denote the 
Schwarzschild radial coordinate and Schwarzschild time. The surface $R
= 2M$ is the event horizon, $R = 0$ is a physical singularity, and $R
\rightarrow \infty$ is spatial infinity.

\begin{figure}[t]
\centering
\includegraphics[width=60mm]{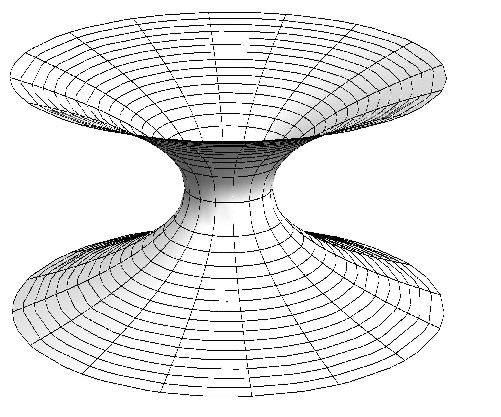}
\caption{Embedding diagram of a two-dimensional slice ($T =
  \textrm{const.}, \theta = \pi/2$) of the extended Schwarzschild
  solution. The distance to the rotation axis is $R$. A wormhole with
  a throat at $R = 2M$ connects two asymptotically flat ends.} 
\label{fig:emb_wormhole}
\end{figure}

If we make the coordinate transformation $R = \psi^2 r$, where
\begin{equation} 
\psi = 1 + \frac{M}{2r},
\end{equation} the Schwarzschild metric becomes \begin{equation}
 ds^2 = -\left( \frac{1 - \frac{M}{2r}}{1 + \frac{M}{2r}} \right)^2 dT^2 +
\psi^4 ( dr^2 + r^2 d \Omega^2  ). \label{eqn:isotropicSchw} 
 \end{equation} 
These are called (quasi or spatially) isotropic coordinates. 
Topologically, the constant-$T$ slices are 
$\mathbb R_+\times \mathbb S^2 \simeq \mathbb R^3 \setminus \{ 0 \}$, 
while geometrically (measuring proper areas or the Schwarzschild
radius $R$) the slices are wormholes.
The isotropic $r$ does not reach the physical singularity at
$R=0$. For large $r$ we see that $R \rightarrow\infty$, but for small
$r$ we see that once again $R \rightarrow \infty$. There is a minimum
of $R = 2M$ at $r = M/2$. We now have two copies of the space outside
the event horizon, $R > 2M$, and the two spaces are connected by a
wormhole with a throat at $R = 2M$
(Figure~\ref{fig:emb_wormhole}). This wormhole picture of a black hole
forms the basis of the initial data used in current moving-puncture
black-hole simulations.
The point $r = 0$, which represents the second asymptotically flat
end, is referred to as the puncture.

In terms of the conformal 3+1 quantities introduced earlier, the
Schwarzschild metric in isotropic coordinates is \begin{eqnarray}
\tilde{\gamma}_{ij} & = & \delta_{ij},  \label{eqn:isogamma} \\
\psi & = & 1 + \frac{M}{2r}, \\
\tilde{A}_{ij} & = & 0, \\
K & = & 0. \label{eqn:isoK}
\end{eqnarray}  The lapse and shift are \begin{eqnarray}
\alpha & = & \frac{1 - \frac{M}{2r}}{1 + \frac{M}{2r}}, \label{eqn:isoalpha}  \\
\beta^i & = & 0. \label{eqn:isobeta}
\end{eqnarray} If we choose (\ref{eqn:isogamma}) -- (\ref{eqn:isoK}) as our 
initial data, and propagate the data with the lapse (\ref{eqn:isoalpha}) 
and shift (\ref{eqn:isobeta}), then the data will remain unchanged: they
are time-independent data. 

As trivial as the time development of these data is, it is difficult
to reproduce numerically in a standard 3D black-hole evolution
code. The reason is that most codes are not stable when the lapse is
negative, which it is here for $r < M/2$. In a numerical code we
prefer to use a lapse that is always positive, or at least
non-negative. Unfortunately, it is not possible to construct a
time-independent, maximal slice of Schwarzschild with two
asymptotically flat ends and an everywhere non-negative lapse, no
matter what shift conditions
we employ~\cite{Hannam:2003tv,Reimann:2003zd,Reimann:2004pn1}. 
Put another way, a maximal- or 1+log-slicing evolution with two
asymptotically flat ends and with a non-negative lapse cannot reach a
stationary state. We will now see that the way around these earlier
results is to give up one of the asymptotically flat ends.

\subsection{Trumpet solution --- maximal slicing}
\label{sec:trumpets}

\begin{figure}[t]
\centering
\includegraphics[width=80mm]{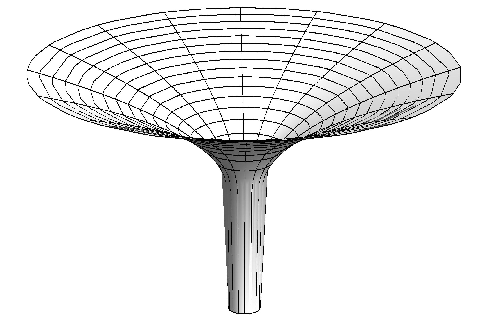}
\caption{Embedding diagram of a two-dimensional slice ($t = \textrm{const.},
  \theta = \pi/2$) of the maximal solution (\ref{eqn:EstaG}) --
  (\ref{eqn:EstaAlp}). The distance to the rotation 
  axis is $R$. In contrast to Figure \ref{fig:emb_wormhole} there is only one
  asymptotically flat end. The other end is an infinitely long cylinder with
  radius $R_0 = 3M/2$.} 
\label{fig:emb_trumpet}
\end{figure}

Since maximal initial data with two asymptotically flat ends cannot be
time independent, we seek an alternative. One alternative is
``trumpet'' data. Consider the maximal slice of the Schwarzschild
spacetime \cite{Reinhardt73,Estabrook73} 
\begin{eqnarray}
\gamma_{RR} & = & \left( 1 - \frac{2M}{R} + \frac{C^2}{R^4}
\right)^{-1} \label{eqn:EstaG}, \\ 
K^i_j & = &   {\rm diag} (2C/R^3, -C/R^3, -C/R^3),  \\
\beta^R & = & \frac{\alpha C}{R^2}, \\
\alpha & = & \sqrt{1 - \frac{2M}{R} + \frac{C^2}{R^4}},  \label{eqn:EstaAlp} 
\end{eqnarray}
with $C = 3\sqrt{3}M^2/4$ and $R \in [1.5M,\infty)$. 
We see that $\alpha \geq 0$ for this domain, and at $R_0 = 3M/2$ the
lapse goes to zero, as does $\beta^R$, while the spatial metric
diverges as $R \rightarrow R_0$, and so the proper distance from $R_0$
to any $R > R_0$ is infinite. As we approach $R_0$ the timeslice
becomes an infinitely long cylinder of radius $3M/2$. This behaviour 
is illustrated in Figure~\ref{fig:emb_trumpet} and we will refer to
these data as ``trumpets''. These data, like (\ref{eqn:isogamma}) --
(\ref{eqn:isoK}) with (\ref{eqn:isoalpha}) and (\ref{eqn:isobeta}),
are time independent.  

With the benefit of hindsight we see that, rather than adopting the
wormhole puncture data (\ref{eqn:isogamma}) -- (\ref{eqn:isoK}) for
numerical evolutions, we might be better off transforming
(\ref{eqn:EstaG}) -- (\ref{eqn:EstaAlp}) to isotropic-like coordinates
(such that $r = 0$ corresponds to $R = 3M/2$), and using those data
instead. Such a transformation was calculated numerically in
\cite{Hannam:2006xw}, and it was shown that these data can indeed be
evolved stably, and are time-independent up to (convergent) numerical
errors. An analytic transformation to isotropic coordinates was given
in \cite{Baumgarte:2007ht}.

\subsection{Trumpet solution --- 1+log slicing}
\label{sec:trumpets2}

The data (\ref{eqn:EstaG}) -- (\ref{eqn:EstaAlp}) are maximally
sliced, $K = 0$. In a numerical simulation, we must solve an elliptic
equation at each time step to find the appropriate lapse function that
maintains maximal slicing (for these data we could assume that the
lapse remains constant, but this will not be true in general
black-hole simulations). This is computationally expensive (i.e.,
slow), and it is therefore currently more practical to
choose a slicing condition so that the lapse can be calculated from an
evolution equation in the same way as all of the other dynamical variables.
One such slicing condition is
\beq
\partial_t \alpha = - n \alpha K \label{eqn:old1plog},
\eeq where $n$ is some constant, usually chosen to be $n = 2$. We see
that if $K = 0$, then the lapse does not evolve, and this condition will
maintain maximal slicing for our trumpet data. This slicing condition
is part of a class of conditions called ``1+log'' slicing \cite{Bona1995}. 
The condition
(\ref{eqn:old1plog}) is actually not a ``geometric'' slicing condition 
in the sense that the slicing resulting from this condition also
depends on the shift $\beta^i$, i.e., on the spatial coordinates
\cite{Bona1995,Alcubierre2003:hyperbolic-slicing}. Furthermore, in
binary simulations it has been found that unphysical gauge modes arise
when (\ref{eqn:old1plog}) is used \cite{vanMeter:2006vi}, and a preferred
choice is 
\beq
(\partial_t - \beta^i \partial_i) \alpha = - n \alpha K. \label{eqn:1plogfull} 
\eeq 
This is equivalent to 
\beq
\lie_{\hat{n}} \alpha = - n K,
\eeq where $\hat{n}$ is the unit timelike normal to the slice and
$\lie$ is the Lie derivative. For the rest of this article, whenever
we refer to 1+log slicing, we will mean (\ref{eqn:1plogfull}). We will
refer to Eqn.~(\ref{eqn:old1plog}) as ``asymptotically maximal
slicing'' because, if it leads to a time-independent geometry, then
that geometry will be maximally sliced.

Since the slicing condition (\ref{eqn:1plogfull}) has proven rather
beneficial in black-hole simulations, we would like to know what the
corresponding time-independent Schwarzschild data are. A stationary
1+log solution was found in Paper I. 
An earlier result on stationary 1+log slices can be found
in~\cite{Buchman:2005ub}, however its exact relation to the present
discussion of punctures remains to be understood.
Here we will present one (of many possible) derivations of the
solution of Paper I.

\subsubsection{Height function derivation}

We begin with the Schwarzschild metric in Schwarzschild coordinates,
Eqn.~(\ref{eqn:schwarzschild}), 
and introduce a new time variable, $t$, related
to Schwarzschild time by a spherically symmetric height function,
$h(R)$, as in \cite{Malec:2003dq},
\beq
  t = T - h(R). \label{eq:timetransf}
\eeq The lapse, shift and future-pointing normal vector in the new
coordinates are \bea
\alpha & = & \sqrt{ \frac{f}{ 1 - f^2 h'^2 } }, \\
\beta^R & = & \frac{ f^2 h'}{f^2 h'^2 - 1}, \\
n_{\mu} & = & \left( - \alpha, 0, 0, 0 \right).
\eea 
Here again $f=1-2M/R$, and we have introduced the notation $h' = \partial 
h(R) / \partial R$. We note that only the derivative of the height
function appears in the new metric.

It is also useful to note that for any stationary spherically
symmetric spacetime the extrinsic curvature is given by
\begin{eqnarray}
K_{RR} & = & \frac{\beta'}{\alpha^2},  \label{eqn:KRR} \\
K_{\theta \theta} & = & R \beta, \\
K_{\phi \phi} & = & R \beta \sin^2 \theta, \label{eqn:KPP}
\end{eqnarray} where 
$\beta = \sqrt{\beta_i \beta^i}$ and 
$\beta' = \partial \beta / \partial R$,
and the trace of the extrinsic curvature $K = K_i^i$ is given by 
\beq
K = \frac{2\beta}{R} + \beta'.
\label{eqn:Ktrace}
\eeq 
For the Schwarzschild solution we also have the relation
\beq
\alpha^2 - \beta^2 = 1 - \frac{2M}{R}. \label{eqn:killingreln}
\eeq 

It will simplify the following calculations if we write $h'$ in terms
of the lapse as \beq
h' = - \frac{\sqrt{\alpha^2 - f}}{f \alpha}, \label{eq:hprime}
\eeq and the shift is now written as \beq
\beta^R =  \alpha \sqrt{ \alpha^2 - f }, \label{eqn:1plogshift}
\eeq where we have chosen the sign of the root to give a non-negative shift.

Following \cite{Malec:2003dq}, the trace of the extrinsic curvature, $K$, can
be written as \bea
K & = & \frac{1}{\sqrt{-g}} \left( \sqrt{-g} n^{\mu} \right)_{,\mu} \nonumber \\
& = & \frac{1}{R^2} \partial_R ( R^2 h' f \alpha ). \label{eqn:trK}
\eea In turn, our 1+log gauge condition can be written for a time-independent
solution as \beq 
\beta^R \partial_R \alpha = n \alpha K. \label{eqn:1plog}
\eeq Combining Equations (\ref{eqn:trK}) and (\ref{eqn:1plog}) and solving
for $\alpha'$, we arrive at the following first-order differential
equation for the lapse, \beq
\alpha' = - \frac{ n (3M - 2R + 2 R \alpha^2) }{R (R - 2M + n R \alpha - R
  \alpha^2) }.  \label{eqn:alpprime}
\eeq 
The solution of this equation is an implicit equation for the lapse, 
\beq
\frac{n}{2} \left[ 3 \ln R + \ln( 2M - R + R \alpha^2 ) \right] -
\alpha = {\rm constant}, 
\eeq
or, more conveniently,
\beq
\alpha^2 = 1 - \frac{2M}{R} + \frac{C(n)^2 e^{2 \alpha / n}
}{R^4}. \label{eqn:alpsoln} 
\eeq 

We will now determine $C(n)$. The equation (\ref{eqn:alpprime}) is
singular when the denominator is zero. We search for a $C(n)$ such
that the numerator and denominator are zero at the same point, and the
equation remains regular (the singularity in the equation signifies a
transition between an elliptic and a hyperbolic problem as discussed
in \cite{Garfinkle:2007yt}). The numerator is zero when \beq
\alpha = \pm \sqrt{1 - \frac{3M}{2R}}.
\eeq We want solutions with an everywhere non-negative lapse, so we
choose the positive root. Substituting this into the denominator of
(\ref{eqn:alpprime}) we find that the denominator is zero for a
particular value of $R$,  
\beq
R_c = \frac{ 3 n^2 M + \sqrt{ 4 n^2 M^2 + 9 n^4 M^2 } }{4 n^2},
\eeq 
where we have chosen the positive root to ensure that $R_c$ is always
positive. This allows us to evaluate the lapse at that point as \beq
\alpha^2_c = \frac{\sqrt{4 + 9n^2} - 3n}{\sqrt{4 + 9n^2} +
  3n}. \label{eqn:alpcrit} 
\eeq 
The value of the constant $C(n)$ is now given as \beq
C^2(n) = \frac{ (3 n + \sqrt{ 4 + 9 n^2} )^3 }{128 n^3} e^{-2 \alpha_c
  / n}. \label{eqn:Csoln} 
\eeq  

We now have the 1+log trumpet solution, given by (\ref{eqn:alpsoln}) with
(\ref{eqn:alpcrit}) and (\ref{eqn:Csoln}) for the
lapse, from which we calculate $\beta^R$ from (\ref{eqn:1plogshift}) and
$\gamma_{RR} = 1/\alpha^2$, and the extrinsic curvature is given by
(\ref{eqn:KRR}) -- (\ref{eqn:KPP}). 

Three special cases of the constant $n$ deserve immediate comment. When $n = 2$
we have the form of 1+log slicing commonly used in numerical simulations of
black-hole spacetimes. The results presented in
\cite{Hannam:2006vv,Hannam:2006xw,Brown:2007tb,Garfinkle:2007yt} pertain to
this case, and we have $R_c = 1.54M$, $C^2(n) = 1.55431 M^4$, and
the location of the throat (the smallest real root of
(\ref{eqn:alpsoln}) with $\alpha = 0$) is $R_0 = 1.31241M$. The
horizon is located at $\alpha(R=2M) = 0.376179$, which differs 
by over 20\% from the common ``$\alpha = 0.3$'' rule-of-thumb estimate
of the horizon location.

In the limit $n \rightarrow 0$, Eqn.~(\ref{eqn:1plog}) is degenerate,
and two solutions exist. In one, $\beta^R = 0$, and we recover the
standard Schwarzschild metric in Schwarzschild coordinates. In the
other, $\partial_R \alpha = 0$, and so $\alpha = 1$ everywhere. As we 
approach $n \rightarrow 0$, the radius of the throat $R_0$ approaches
the singularity as $R_0 \sim e^{- 2/(3n)} / (2^{5/3} n)$. 

The third case of interest is the limit $n \rightarrow \infty$. Now
the only physically meaningful solution is that $K = 0$ everywhere,
i.e., maximal slicing. We have $C(\infty) = 3\sqrt{3} M^2 / 4$, and
the radius $R_c$ and the throat coincide at $R_c = R_0
= 3M/2$. This is the cylindrical maximal slice (\ref{eqn:EstaG}) -- 
(\ref{eqn:EstaAlp}).  

\begin{figure}[t]
\centering
\includegraphics[width=60mm]{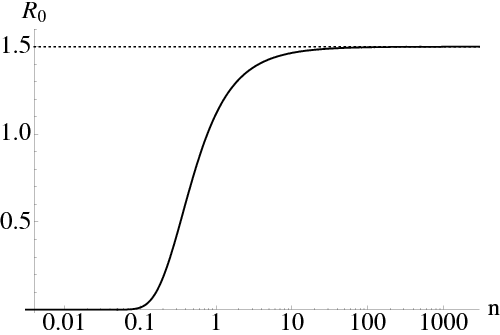}
\caption{Location of the throat, $R_0$, as a function of the coefficient $n$
in the slicing equation (\ref{eqn:1plog}). In the limit $n \rightarrow \infty$
the stationary slice is maximal and the throat is at $R_0 = 1.5M$, indicated
in the figure by a horizontal line. For small values of $n$, the throat
approaches the singularity. }
\label{fig:throatlocation}
\end{figure}

\subsubsection{Penrose diagrams}

One advantage of working in spherical symmetry is that redundant 
coordinates may be suppressed and we can visualize the way the spacetime 
is sliced on two-dimensional diagrams, such as the Carter-Penrose diagram.

In order to do this for a given solution (\ref{eqn:alpsoln}), we first
integrate the height function $h$ using (\ref{eq:hprime}) to obtain
$T$ for every $R$. The singularity at $R = 2M$ can be handled (at
least numerically) by introducing a different quantity, such as
$e^{-h}$, around the horizon. The undetermined value of $t$ in
Eqn.~(\ref{eq:timetransf}), which can be interpreted as the constant
of integration, expresses the fact that we do not calculate a single
slice but a foliation of the Schwarzschild spacetime. As expected, the
slices are related to each other by sliding along the Killing vector
field $\partial_T$. 

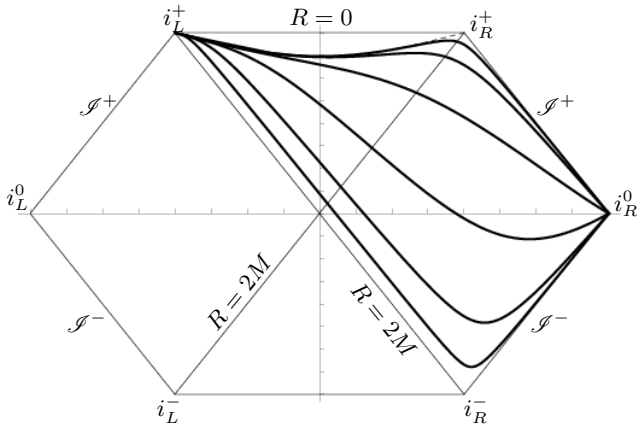
\begin{figure}[t]
\centering
\begin{overpic}[width=80mm]{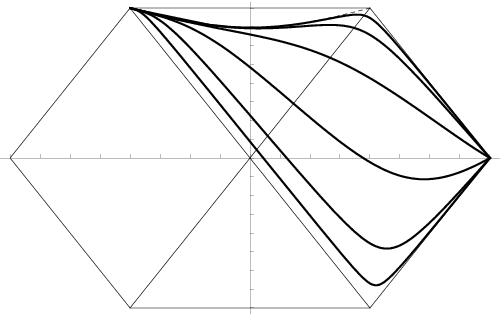}
\put(75,62){$i_R^+$}
\put(24,63){$i_L^+$}
\put(86,48){$\mathscr{I}^+$}
\put(10,48){$\mathscr{I}^+$}
\put(8,12){$\mathscr{I}^-$}
\put(85,12){$\mathscr{I}^-$}
\put(74,-2){$i^-_R$}
\put(23,-2){$i^-_L$}
\put(-2,33){$i_L^0$}
\put(99,32){$i_R^0$}
\put(33,12){\begin{rotate}{51}
  $R = 2M$
\end{rotate}}
\put(55,19){\begin{rotate}{-51}
  $R = 2M$
\end{rotate}}
\put(45,63){$R = 0$}
\end{overpic}
\caption{The Penrose diagram of the slices defined by the stationary
  solution of the 1+log condition with $n = 2$. Every slice approaches
  $i_L^+$ along the curve $R = R_0 \approx 1.31 M$ and spatial
  infinity $i_R^0$ along a curve of constant $T$. The slices are
  displayed in time steps of $4M$.} 
\label{fig:Penanalyt}
\end{figure}

From the coordinates $(R,T)$ along one slice we transform to Kruskal coordinates 
$(u,v)$ by either
\bea
 u &=&  \sqrt{\frac{R}{2M} - 1} \: e^{R/(4M)} \cosh \frac{T}{4M} \label{eq:Kruskalu},\\
 v &=&  \sqrt{\frac{R}{2M} - 1} \: e^{R/(4M)} \sinh \frac{T}{4M}\label{eq:Kruskalv},
\eea for $R > 2M$ or
\bea
 u &=&  \sqrt{1 - \frac{R}{2M} } \: e^{R/(4M)} \sinh \frac{T}{4M} \label{eq:Kruskalu2}, \\
 v &=&  \sqrt{1 - \frac{R}{2M} } \: e^{R/(4M)} \cosh \frac{T}{4M},
\eea for $R < 2M$. 
Compactifying the result via 
\beq
 u \pm v = \tan (U \pm V),
\eeq
where $U$ and $V$ are the abscissa and the ordinate of the Penrose
diagram, yields the picture displayed in  
Figure~ \ref{fig:Penanalyt}. 
Note that we chose non-negative shift $\beta^R$, which also 
determines the sign of $h$ (as well as $u$ and $v$ in
Eqns.\ (\ref{eq:Kruskalu}) and (\ref{eq:Kruskalv})) . The opposite
choice would lead to slices that are mirror images of those in
Figure~\ref{fig:Penanalyt}, connecting $i_R^+$ to $i_L^0$. 

\subsubsection{Trumpet data in isotropic coordinates}
\label{sec:isotrumpet}

We have now derived the stationary 1+log solution. In Paper I we
compared this solution with the late-time data from moving-puncture
evolutions of wormhole puncture data, which we will discuss again in
Section~\ref{sec:numerics}. We would also like to put this solution
into isotropic-like coordinates, as was done for the maximal trumpet
data in \cite{Hannam:2006xw,Baumgarte:2007ht}. This provides a good
test-case for numerical evolution codes, and could be a starting point
for generating trumpet data for black-hole binaries. 

The implicit nature of the solution (\ref{eqn:alpsoln}) makes it
difficult to analytically construct the transformation to isotropic
coordinates. However, solving (\ref{eqn:alpsoln}) for a function
$R(\alpha)$ leads to four roots of a fourth order polynomial, which,
if chosen appropriately, represent the analytical solution.
In this section, $R$ should always be understood as this function of
$\alpha$, whereas $\alpha$ becomes the 
independent variable that parametrizes the spatial dependency.
Apart from that our approach is similar to the one
used in \cite{Baumgarte:2007ht}. We note that the
$\gamma_{RR}$ component of the stationary 1+log metric can be related to 
the $\gamma_{rr}$ in isotropic coordinates by \begin{eqnarray}
\gamma_{RR} & = & \left( \frac{\partial r}{\partial R} \right)^2 \gamma_{rr} \\
& = & \left( \frac{\partial r}{\partial R} \right)^2 \psi^4.
\end{eqnarray} We therefore find that, using $R = \psi^2 r$ and
$\gamma_{RR} = \alpha^{-2}$, 
\beq
\frac{\partial r}{\partial R} = \frac{r}{\alpha R}, 
\eeq and a relation between the isotropic coordinate $r$ and
Schwarzschild $R$ may be found by either \begin{eqnarray}
r & = & \exp \left( \int^\alpha \frac{1}{\bar \alpha R} \frac{dR}{d
  \alpha}(\bar \alpha) \: d \bar \alpha  \right), \ \ \ \ \ {\rm
  or} \label{eqn:rtoR1}\\  
r & = & R^{1/\alpha} \exp \left( - \int_\alpha^1 \frac{\ln R}{\bar
  \alpha^2} d\bar \alpha  
\right), \label{eqn:rtoR2} 
\end{eqnarray} where the last equation is obtained by integration by
parts and the upper integration limit is chosen such that $r
\rightarrow R$ as $\alpha 
\rightarrow 1$, i.e., towards  
spatial infinity. Both integrals (\ref{eqn:rtoR1}) and (\ref{eqn:rtoR2})
diverge as $\alpha \rightarrow 0$, 
but (\ref{eqn:rtoR1}) diverges less strongly, and can be integrated
numerically to arbitrarily small $\alpha$ 
with sufficient accuracy to produce data suitable for a numerical
evolution. On the other hand, (\ref{eqn:rtoR2}) has the attractive
property that as $\alpha \rightarrow 1$ 
the factor $R^{1/\alpha}$ 
gives the asymptotic behaviour that we wish. In practice, we choose a
point $\alpha_s = 0.1$, and for $\alpha < \alpha_s$ use \beq
r(\alpha) = R(\alpha_s)^{(1/\alpha_s)} \exp \left[ -
 \int_{\alpha}^{\alpha_s} \frac{1}{\bar{\alpha} R(\bar{\alpha})}
 \frac{dR}{d\alpha}(\bar{\alpha}) \: d\bar{\alpha} - C_0 \right], 
 \eeq where \beq
 C_0 = \int_{\alpha_s}^{1} \frac{\ln R(\alpha)}{\alpha^2} d\alpha.
 \eeq For $\alpha > \alpha_s$ we use \beq
 r(\alpha) = R(\alpha)^{(1/\alpha)} \exp \left[ 
 \int_{\alpha_s}^{\alpha} \frac{\ln R(\bar{\alpha})}{\bar{\alpha}^2}
 d\bar{\alpha} - C_0 \right]. \eeq 
Once we have the solutions 
$R(\alpha)$ and $r(\alpha)$, we may construct $r(R)$ to 
 whatever accuracy is desired, and then transform our data to isotropic
 coordinates via \begin{eqnarray} 
 \psi & = & \sqrt{ \frac{R}{r} }  \label{eqn:Tpsi}, \\
 \beta^r & = & \frac{\partial r}{\partial R} \beta^R ,\\
 K_{rr} & = & \left( \frac{\partial r}{\partial R} \right)^{-2}
 K_{RR}. \label{eqn:TK} 
 \end{eqnarray} 

Note that the singularity in the conformal factor is now milder than
in the standard puncture case, where $\psi \sim 1/r$, while here
\beq
	\psi(r) = \sqrt{\frac{R(\alpha(r))}{r}} \simeq
                  \sqrt{\frac{R_0}{r}} \sim \frac{1}{\sqrt{r}}. 
\eeq
This fact is intuitively expressed in the embedding diagrams in
Figures~\ref{fig:emb_wormhole} and \ref{fig:emb_trumpet}: the
expansion of the wormhole geometry as compared to the trumpet geometry
leads to a stronger singularity of the conformal factor. Since
wormholes and trumpets both allow a representation on $\mathbb R^3$
where a coordinate singularity at $r=0$ is absorbed in the conformal
factor $\psi$, we refer to both cases as punctures.

\begin{figure}[t]
\centering
\includegraphics[width=80mm]{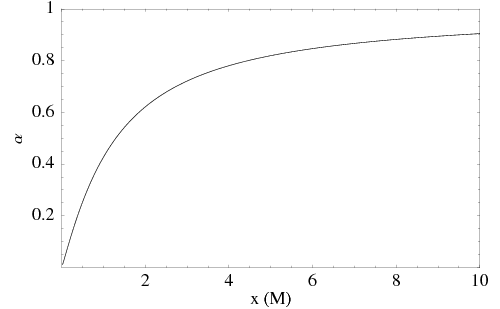}
\includegraphics[width=80mm]{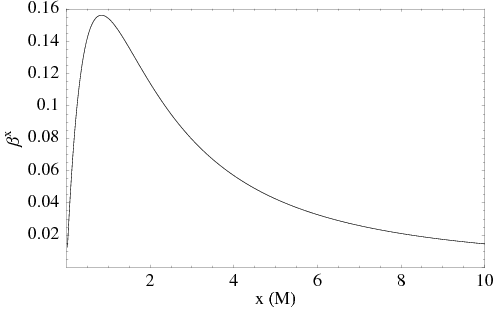}
\includegraphics[width=80mm]{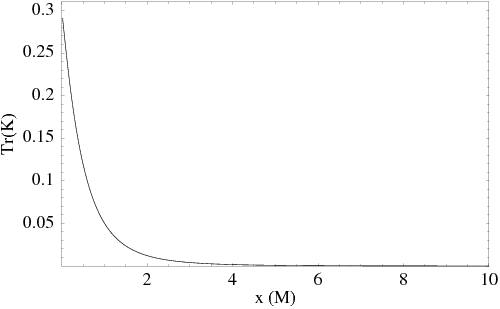}
\caption{The lapse $\alpha$,  $x$-component of the shift vector $\beta^x$,
 and $K$ for time-independent 1+log data in isotropic coordinates. The
 data are shown along the $x$-axis.}
\label{fig:StatData}
\end{figure}

All of the numerical calculations described here were performed with
Mathematica, and the data output as tables of physical quantities
parametrized by the isotropic coordinate $r$. The data files were then
read into our full 3D code {\cite{Bruegmann:2003aw,Bruegmann:2006at}},
where they were transformed to Cartesian coordinates and interpolated
onto the numerical grid. Examples of the resulting data for $\alpha$,
$\beta^x$ and $K$ are shown in Figure~\ref{fig:StatData}. The time
independence of these data will be explicitly demonstrated in 
numerical evolutions in Section~\ref{sec:trumpetevolution}. 

\section{Numerical simulations}
\label{sec:numerics}

We now turn to the numerical evolution of wormhole and trumpet
puncture data for the Schwarzschild spacetime. We start with a brief
description of the moving-puncture method and our numerical
techniques, as well as a summary of our procedure for analyzing our
results (including the construction of Penrose diagrams), before
finally presenting our numerical results.

\subsection{The BSSN/moving-puncture system} 

The 3+1 decomposition provides evolution equations for the spatial metric
$\gamma_{ij}$ and extrinsic curvature $K_{ij}$ \cite{York79}. The BSSN
reformulation consists of rewriting the evolution equations in terms of
conformally rescaled variables, $\{\psi, \tilde{\gamma}_{ij},
\tilde{A}_{ij}, K\}$, where we now use $p = 4$ in (\ref{eqn:Adecomp}),
and an additional variable is introduced: $\tilde{\Gamma}^i = -
\partial_j \tilde{\gamma}^{ij}$.  

When we deal with puncture data, the conformal factor $\psi$ diverges
at each puncture, and this is handled in the moving-puncture
modification {\cite{Campanelli:2005dd,Baker05a}} of the BSSN system by
replacing $\psi$ with either the variable $\phi = \ln \psi$ or
$\chi=\psi^{-4}$ (or $\chi = \psi^{-2}$ \cite{Marronetti:2007wz}), and
one of these quantities is evolved instead. The details of the BSSN
system are given in \cite{Shibata95,Baumgarte99}. We use the BAM code,
and provide details of our implementation of the BSSN/moving-puncture
system in~\cite{Bruegmann:2006at}.

Given evolution equations for the variables $\{ \tilde{\gamma}_{ij},
\tilde{A}_{ij}, K, \tilde{\Gamma}^i \}$ and $\phi$ or $\chi$, and some
initial data, we also need to choose a lapse 
and shift during the evolution. We have already discussed the 1+log
evolution equations (\ref{eqn:old1plog}) and (\ref{eqn:1plogfull}) for
the lapse function; the stationary slices for these slicing conditions
are given in Section~\ref{sec:trumpets2}. For the shift vector we use
the $\tilde{\Gamma}$-driver condition \cite{Alcubierre01a,Alcubierre02a}, 
\begin{eqnarray}
\partial_t \beta^i & = & \frac{3}{4} B^i,  \label{eqn:driver1} \\
\partial_t B^i & = & \partial_t \tilde{\Gamma}^i - \eta B^i. 
\label{eqn:driver2}
\end{eqnarray} 
This shift condition is crucial to the behaviour of the
moving-puncture system.  In black-hole binary simulations it generates
a shift that moves the punctures around the grid on trajectories that
match very well the motion that would be seen from infinity
\cite{Hannam:2006vv,Hannam:2007ik}. This shift {\em also} allows the
wormhole puncture data to evolve to the stationary 1+log geometry, as
shown in Paper I, and as we will see again in Section~\ref{sec:numresults}.  
As we said in the Introduction, the approach to the ``puncture geometry''
requires a dramatic stretching of the coordinate representation of the
slices. We will illustrate this extreme behaviour in detail in the coming
sections, but it can be seen most immediately by calculating the norm of the
shift vector, $\beta^2 = \gamma_{ij} \beta^i \beta^j$, during the first few
$M$ of evolution: although $\beta^i$ remains finite, and in fact goes
to zero as we approach the puncture, $\beta^2$ diverges. This can also
be seen analytically \cite{Tuite08}.
The wormhole slice is stretched such that all of the numerical points
extremely quickly leave the part of the slice that cannot be
stationary, and the part of the slice in the second copy of the
exterior space, and the points relax onto the stationary 1+log
slice. This could not happen with a zero shift, and of course would
not happen with an arbitrary shift. There may be a large class of
shift conditions that produce the same effect, but the first to be
found, and the one that is standard in moving-puncture simulations, is
the $\tilde{\Gamma}$-driver condition (\ref{eqn:driver1}) --
(\ref{eqn:driver2}), and variants.

\subsection{Penrose diagrams from numerical data} 
\label{sec:numpenrose}

A convenient way to view the numerical evolution of our data is to
represent them in a Penrose diagram. Given the Schwarzschild
coordinates $R$ and $T$ of a given point, it is straightforward to
calculate the corresponding point on a Penrose diagram. From our
numerical data we can easily calculate the Schwarzschild coordinate,
$R(r,t)$; see Section~\ref{sec:numresults} and
Eqn.~\ref{eqn:Rcartesian}. Each point on the initial
slice is at a constant Schwarzschild time $T(r,0) = T_0$, which we are
free to choose, but $T(r,t)$ is not known. Although a variety of ways
to compute an appropriate coordinate 
representation suggest themselves, it actually turned out to be
somewhat tricky to find one that works reliably with our numerical
data, see also the discussion in {\cite{Brown:2007nt}} where the
transformation to Kruskal coordinates is implemented. Care has to be
taken near the horizon at $R=2M$, and in our case some issues arose at
mesh refinement boundaries.  

\begin{figure}[t]
\centering
\begin{overpic}[width=80mm]{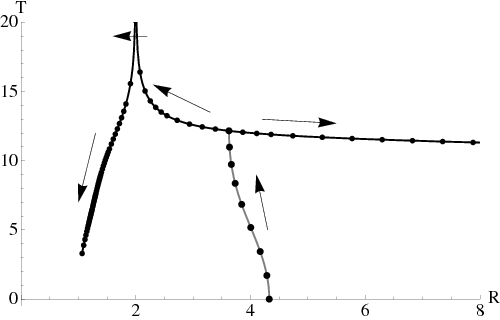}
\put(54,22){$\int \dot T dt$}
\put(56,42){$\int T' dr$}
\put(35,46.5){$\int T' dr$}
\put(24,61){$\int u' dr$}
\put(5.5,31){$\int T' dr$}
\put(43,25){\begin{rotate}{-70}
$r=\rm const.$
\end{rotate}}
\put(82,31){\begin{rotate}{-2}
$t=\rm const.$
\end{rotate}}
\end{overpic}
\caption{A sketch of the method we use to obtain $T(r,t)$. The lines
  are drawn from actual data with $r_0 = 3.25 M$ for the integration
  in time and $t = 10M$ for the integration in space. Only a subset of
  the grid points is displayed.} 
\label{fig:Int_method}
\end{figure}
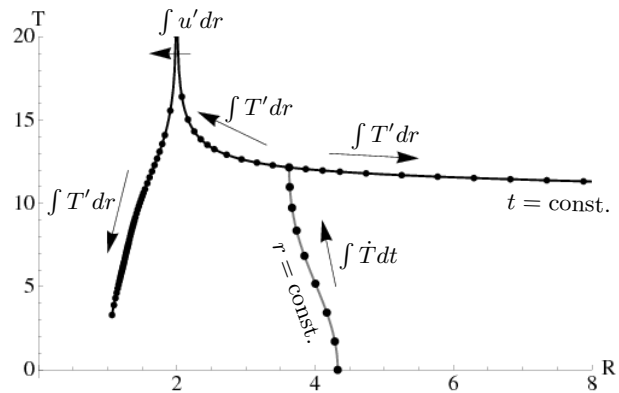

An outline of the procedure that we settled on is as follows. 
The method is also illustrated in Figure~\ref{fig:Int_method}.
We first choose a numerical point $r_0$ far from the black hole (in
practice a numerical point that does not pass through the horizon 
during the evolution). As we have already said, on the initial slice
that point has Schwarzschild time $T(r_0,0) = T_0$. We use a
differential equation for $\dot{T} = \partial T/  \partial t$ to
integrate forward in time through our numerical data and produce
$T(r_0,t)$. Then, on each time slice, $t = t_i$, we use a second
differential equation, this time for $T' = \partial T / \partial r$,
to integrate across the slice and calculate $T(r,t_i)$. The equation
we use is badly behaved near the horizon, and for a set of five points
across the horizon we integrate instead a differential equation for
$u' = \partial u / \partial r$, where $u$ is the Kruskal coordinate
defined by (\ref{eq:Kruskalu}) and (\ref{eq:Kruskalu2}).

We derive the required differential equations by considering the
transformation of the spacetime metric between Schwarzschild and
numerical coordinates. A differential equation 
for $\dot{T}$ can be found using the transformation \begin{equation}
g_{tt} = \left( \frac{\partial T}{\partial t} \right)^2 g_{TT} + \left(
  \frac{\partial R}{\partial t} \right)^2 g_{RR}. 
\end{equation} We therefore find \beq
\dot{T} = \left(1 - \frac{2M}{R} \right)^{\hskip-3pt -1} \sqrt{ {\dot R}^2 -
  \left(1 - \frac{2M}{R} \right) g_{tt} } ~.
\label{eqn:Tdot}
\eeq This expression is valid for $R > 2M$ and $u > 0$, and we choose
$r_0 = 3.25M$, where this is always true. The metric component
$g_{tt}$ is given by $g_{tt} = - (\alpha^2 - \beta_i \beta^i)$.  
 Using (\ref{eqn:Tdot}) we can integrate the Schwarzschild time along
 the constant $r_0 = 3.25M$ curve for the duration of the
 simulation. For the numerical coordinate $r_0 = 3.25M$ we now know
 $R$ and $T$ throughout the numerical evolution.  

A differential equation for $T'$ on a slice of constant numerical coordinate
time can be found by transforming the spatial part of the metric,
\begin{eqnarray}
 \gamma_{rr} &=& \left( \frac{\partial T}{\partial r} \right)^{\hskip-3pt 2}
 g_{TT} + \left( \frac{\partial R}{\partial r} \right)^{\hskip-3pt 2} g_{RR}
 \label{eq:gxxtransform}\\ 
\Rightarrow ~~ {T'}^2 &=& \frac{\gamma_{rr} - {R'}^2 g_{RR}}{g_{RR}} \\
& = & \left(1 - \frac{2M}{R} \right)^{\hskip-3pt -2} \left[ {R'}^2 - \left(1 -
    \frac{2M}{R} \right) \gamma_{rr} \right] ~~. \label{eqn:dTdx2} 
\end{eqnarray} Note that $R$ and $R'$ can be readily computed from the
numerical data, since we are dealing with data on one numerical
timeslice. In order to integrate (\ref{eqn:dTdx2}), we must take a
square root and choose the sign such that the slices go smoothly
through the horizon, and to define which ends of the computational
domain belong to which side of the Penrose diagram. The result is 
\begin{equation}
 T' = - \left(1 - \frac{2M}{R} \right)^{\hskip-3pt -1} \sqrt{ {R'}^2 - \left(1
     - \frac{2M}{R} \right) \gamma_{rr} }~~. \label{eqn:dTdx} 
\end{equation} Equation (\ref{eqn:dTdx}) is integrated along the
entire numerical slice, except at the points near the horizon, where
$T'$ is singular. We overcome this difficulty by instead integrating
the Kruskal coordinate $u$ through five points that cross the
horizon. A differential equation for $u'$ can be found by transforming
the $(r,t)$ coordinates to $(u,R)$: 
\begin{equation}
 \gamma_{rr} =  \left( \frac{\partial u}{\partial r} \right)^{\hskip-3pt 2}
 g_{u u} + 2 \frac{\partial u}{\partial r} \frac{\partial R}{\partial r} g_{u
   R} + \left( \frac{\partial R}{\partial r} \right)^{\hskip-3pt 2} g_{RR}. 
 \label{eqn:gxxK}
 \end{equation} The metric components $g_{uu}$, $g_{u R}$ and $g_{RR}$
can be written in terms of only $u$ and $R$ by making use of the
Schwarzschild metric in Kruskal coordinates, 
 $g_{\mu \nu}^K$:
 \begin{eqnarray}
 v &=& \sqrt{u^2 - \left(\frac{R}{2M}-1 \right) e^{R/(2M)}} \qquad (v>0), \label{eqn:vofu} \\ 
g_{uu} &=& g_{uu}^{K} + \left( \frac{\partial v}{\partial
    u}\right)^{\hskip-3pt 2} g_{vv}^{K} = \left( 1 - \frac{u^2}{v^2} \right)
\frac{32M^3}{R} e^{-R/(2M)} ,\nonumber \\ 
g_{u R} &=& \frac{\partial v}{\partial u}\frac{\partial v}{\partial
  R}g_{vv}^{K} = \frac{4u}{v^2} \frac 1M , \nonumber\\ 
g_{RR} &=& \left( \frac{\partial v}{\partial R}\right)^{\hskip-3pt 2}
g_{vv}^{K} = - \frac{R}{2M} \frac 1{v^2} e^{R/(2M)} ~. 
\end{eqnarray} Producing an equation for $u'$ from (\ref{eqn:gxxK}) once again
requires the appropriate choice of sign for a square root. The final result is 
\begin{equation}
 u' = \frac{1}{g_{uu}} \left( - g_{u R} R' + \sqrt{\left( g_{u R} R' \right)^2
     + g_{uu} \left( \gamma_{rr} - {R'}^2 g_{RR} \right) }
 \right)~~. \label{eqn:dudx} 
\end{equation}
Once equations (\ref{eqn:dTdx}) and (\ref{eqn:dudx}) have been integrated
along a slice, a constant of integration is chosen to give the value already
calculated at $r_0 = 3.25M$ using (\ref{eqn:Tdot}), and a choice of
$T_0$ for the time on the initial slice.

\section{The numerical transition from a wormhole to a trumpet}
\label{sec:numresults}

\subsection{Setup}

Simulations were performed with computational grids consisting of
eight nested cubical boxes. The boxes are labeled by $l = 0, 1, \dots,
7$. Each box contains $N^3$ points, and the grid-spacing in each 
box is $h_l = H/2^l$, where $H$ is the grid-spacing
of the largest box. We denote $h_{min} = h_{l_{max}}$. The simulations
were performed in only one octant of a three-dimensional Cartesian
grid (the data in the other octants being easily deduced  
by exploiting the known spherical symmetry of the Schwarzschild
spacetime). The grid points are staggered across the coordinate axes,
such that the boundaries of the cubical domain are the six planes
defined by $x = h_{min}/2$, $y = h_{min}/2$,  
$z = h_{min}/2$, $x = X$, $y = Y$ and $z = Z$,
with $X = Y = Z = 192M$,  which define the standard ``outer boundary'' of the
 computational domain.
Three simulations were performed to make a convergence series, with $N
= {64,96,128}$, and $H = {6M,4M,3M}$, $h_{min} =
{M/21.33,M/32,M/42.67}$. Note that these grid configurations are the
same as used for the single black-hole tests in \cite{Bruegmann:2006at}. 

The numerical simulations discussed in this section start with the initial
data (\ref{eqn:isogamma}) -- (\ref{eqn:isoK}), and initial shift $\beta^i = 0$
and initial lapse $\alpha = 1$. A number of additional simulations were
performed with initial lapse $\alpha = \psi^{-2}$ for comparison.

The results of the numerical evolution are shown in
Figures~\ref{fig:embedding} to \ref{fig:Pen_late}. The data for these
figures were produced by interpolating numerical data onto the $x$
coordinate axis. The spherical symmetry of the solution allows us to
analyze the simulation using these data only. The Schwarzschild radial
coordinate can be calculated by relating the numerical spatial metric
to the Schwarzschild metric (\ref{eqn:schwarzschild}). We know  
that $\gamma_{\theta \theta} = R^2$, and because of the spherical
symmetry we also have $\gamma_{\theta \theta}  = ( \partial x^i /
\partial \theta ) (\partial x^j / \partial \theta ) 
\gamma_{ij} = ( \partial x^i / \partial \theta ) (\partial x^j / \partial
\theta ) \psi^4 \tilde{\gamma}_{ij}$.  
Along the $x$-axis, $\theta = \pi/2$, and so $\partial x / \partial \theta =
\partial y / \partial \theta = 0$ 
and $\partial z / \partial \theta = x$, and we have \beq
R^2 |_{y=z=0} = \psi^4 x^2 \tilde{\gamma}_{zz}. \label{eqn:Rcartesian}
\eeq

\subsection{Early-time behaviour}

\begin{figure}[t]
\centering
\includegraphics[width=40mm]{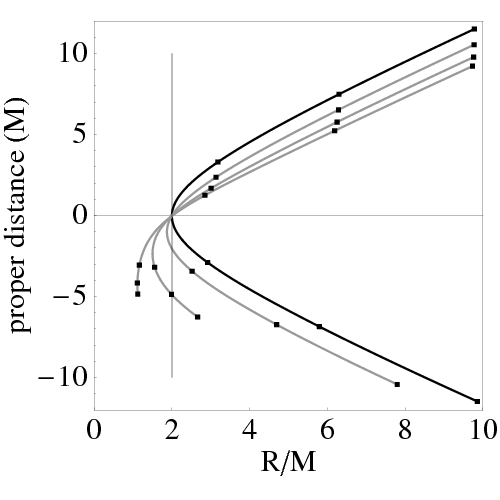}
\quad
\includegraphics[width=40mm]{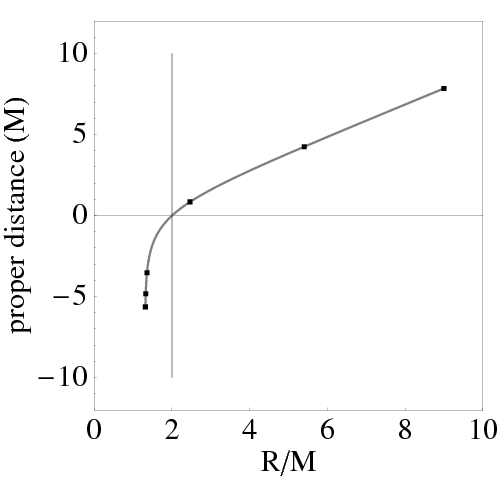}
\includegraphics[width=40mm]{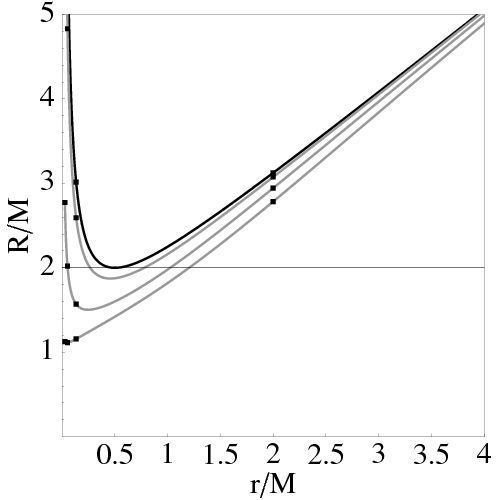}
\quad
\includegraphics[width=40mm]{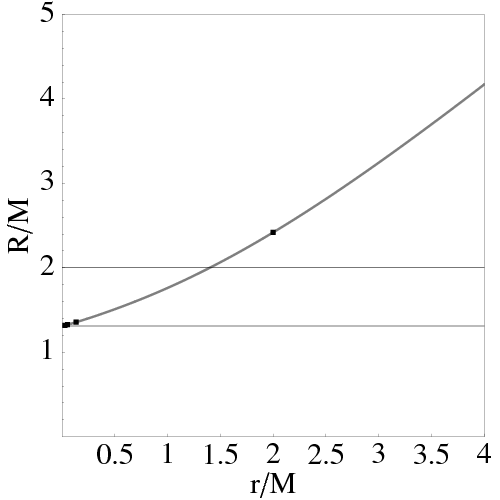}
\caption{The top two figures show the proper separation from the (outer)
  horizon versus the Schwarzschild coordinate $R$. The left panel shows the
  slices at $t = 0,1,2,3M$, and the right panel shows the slice at $t =
  50M$. The final numerical slice terminates at $R \approx 1.31M$. The vertical
  line indicates the horizon at $R = 2M$, and the six points represent $x/M =
  1/40,1/20,1/8,2,5,8$ on each slice. The lower two figures show Schwarzschild
  $R$ versus numerical $r$ for the same points at the same times. The
  horizontal lines show $R = 2M$ and $R = 1.31M$.}
\label{fig:embedding}
\end{figure}

Figure \ref{fig:embedding} illustrates the main global feature of the
time development of the numerical
slices, namely the transition from wormhole to trumpet
asymptotics. The upper figure shows the proper distance of a given point
from the horizon at $R = 2M$ versus that point's Schwarzschild
coordinate $R$. The thick line indicates the $t=0$ data. 
Initially $R = 2M$ is the minimal surface, and the
data contain two copies of the space outside the horizon. This is the
initial wormhole that we have referred to several times. If we
consider a surface of revolution around the proper time axis, we
obtain just another version of the standard picture 
of a spacetime wormhole, as shown earlier in
Figure~\ref{fig:emb_wormhole}. We show the data only up to coordinate
$R = 10M$; the upper and lower lines in the figure quickly asymptote
to $\pm$45~degrees, and do not add extra information to the figure. If
we follow the upper line outwards, we move further from the origin
(the puncture), and the plot ends at $r =8.25M$. If we follow the
lower line outwards, we move closer to the puncture, and in this case
the plot ends at $r \approx M/40$. The corresponding picture of $R$ in
terms of coordinate $r$ is shown in the lower panels of
Figure~\ref{fig:embedding}. 

At early times two notable things happen. First, the minimal 
surface shifts to $R < 2M$, and the numerical domain contains two
surfaces with $R = 2M$, which we will call the ``inner'' and ``outer''
horizons. The proper distance in Figure~\ref{fig:embedding} is with
respect to the outer horizon. Second, the points in the lower right
part of the figure rapidly move to the left. In other words, the
Schwarzschild $R$ corresponding to those points rapidly decreases. The
numerical points don't ``go'' anywhere, of course; they are points on
a fixed grid. But their location in the Schwarzschild spacetime {\it
  does} change, and quickly. Within only $3M$ all of the points have
passed the inner horizon, and it has ceased to be part of the
numerical domain.  

At later times ($t > 40M$) the points close to the puncture settle at
a constant value of $R$, which we will soon see is close to $R_0 =
1.31M$, the location of the throat in the stationary 1+log solution
derived in Section~\ref{sec:trumpets2}. This is shown in the upper right
figure. At first sight this does {\it
  not} correspond to the cylindrical asymptotics shown in
Figure~\ref{fig:emb_trumpet}; the ``cylinder'' looks too short. The
reason is that the spatial metric now diverges more slowly as we
approach the asymptotic region, and so a point initially ``close'' to
the second asymptotically flat end was a larger proper distance from
the outer horizon than it is now that it's ``close'' to the cylinder.  

\begin{figure*}[t]
\centering
\includegraphics[width=160mm]{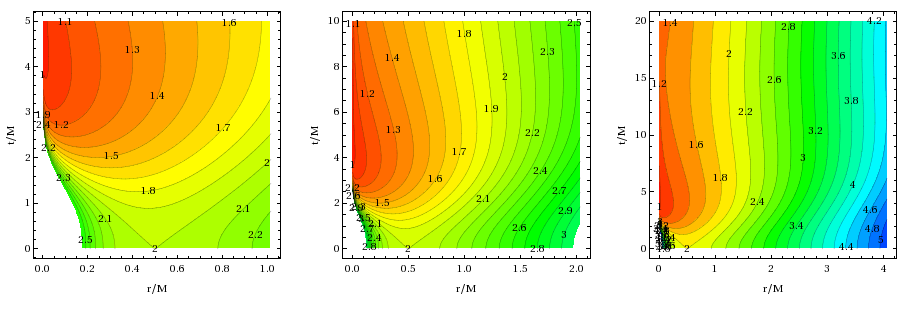}
\caption{
Contour plots showing the behaviour of $R(r,t)$ on the numerical grid
for the first $20M$ of a simulation. 
The axes of the figures give $r$ and $t$ in units of $M$ for three
different coordinate ranges. The contour-labels and the colours
indicate the value of $R$.
We clearly see that $R$ changes rapidly for a given value of the
$r$-coordinate for the first few $M$ of evolution before settling down
to approximately stationary values at later time. Points with $r >
M/2$ accelerate towards the black hole ($R$ decreases) before settling
down again at some smaller value of $R$.
As the slice moves towards the Schwarzschild singularity, the horizon,
initially at $r = M/2$ and $R=2M$, splits into two copies, and values
of $R < 2M$, which are not present in the initial data, appear as time
progresses. The initial slice covers the interior of the black hole
from $R=2M$ to $R=+\infty$ for $r=M/2$ to $r=0$. This region is
quickly squeezed towards zero extent in $r$, as the lines of constant
$R$ in the lower left of the panels indicate. Note that for given $r$,
the coordinate motion $R(r,t)$ is not monotonic in $t$, but $R(r,t)$
approaches its asymptotic value via a damped oscillation, see also
Figure~\ref{fig:RMotionLate}.
}
\label{fig:Rofrt}
\end{figure*}

The lower two panels in Figure~\ref{fig:embedding} give a
complementary picture. These plot Schwarzschild $R$ versus the
numerical coordinate $r$ to illustrate directly how the Schwarzschild
$R$ of each grid point changes.
The behaviour at early times in terms of $R(r,t)$ is also shown in
Figure~\ref{fig:Rofrt}, which shows several contour plots of $R$
(indicated by variations in colour) at different values of coordinate
$r$ and time $t$. These figures illustrate many of the main features
of the early-time evolution of the wormhole puncture data. We can
clearly see that the horizon, initially at $r = M/2$, splits into two
copies, and also that values of $R < 2M$, which are not present in the
initial data, appear as time progresses. We can also see that for
larger $r$ the value of $R$ changes very little, although it does
decrease for all $r$.

\begin{figure}[t]
\centering
\includegraphics[width=60mm]{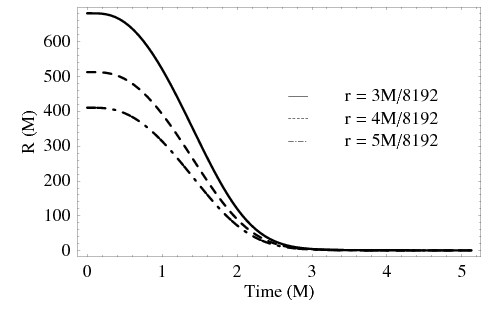}
\includegraphics[width=60mm]{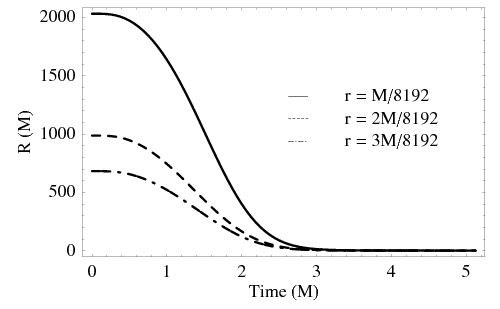}
\caption{Evolution of $R$ for fixed $r$ for the first $5M$ of an
  extremely high-resolution simulation. Top panel: the evolution of
  the third, fourth and fifth closest points to the puncture, at $r =
  \{3,4,5\}M/8192$, are shown. We do not expect the closest two points
  to be reliable, due to finite-difference errors across the puncture,
  although they show similar behaviour, shown in the lower plot.}
\label{fig:Rmotion}
\end{figure}

The behaviour of grid points close to the puncture is shown in Figure~\ref{fig:Rmotion}. 
This plot uses data from a simulation that used extremely high resolution at the 
puncture: 16 nested boxes, each containing $64^3$ points, with a coarsest resolution
of $H = 4M$ and a finest resolution of $h_{min} = M/8192$; the simulation was run for 
$t = 5M$, in order to obtain the results displayed in Figure~\ref{fig:Rmotion}.

As shown in Paper I,
the numerical data become discontinuous
across the puncture. This means that finite-difference derivatives (which are
used to calculate many quantities in the BSSN evolution) will have even worse
discontinuities, and the numerical method cannot converge for points that are
within a stencil-width of the puncture, which for the simulations discussed
here includes the two grid points closest to the puncture. Fortuitously the
nature of the BSSN/moving-puncture system is such that these errors do not
seem to propagate out from the puncture, and clean convergence can be seen up
to the last few grid points. (This is shown in Figures 2 and 3 of
\cite{Bruegmann:2006at}.) In the upper panel of Figure~\ref{fig:Rmotion} we
show the time development of $R$ for the third, fourth, and fifth closest
points to the puncture of our extremely high-resolution simulation, at $r =
\{3,4,5\}M/8192$. Although the points closer to the puncture are not expected
to show clean convergence, we see in the lower panel of
Figure~\ref{fig:Rmotion} that they display similar behaviour.  

Initially the points in the upper panel are at $R = (1 + 2M/r)^2 r =
\{684,513,410\}M$. The value does not change significantly for the first $M/2$
of the simulation, but then quickly deceases.

\begin{figure}[t]
\centering
\includegraphics[width=60mm]{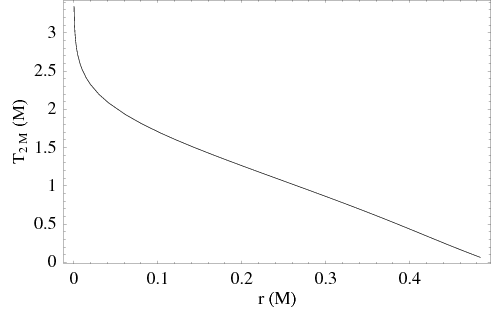}
\caption{Time for grid points to pass the inner horizon, as a function of coordinate $r$.
The innermost grid point is initially at $R = 2031M$, and reaches the inner horizon in
$T_{2M} = 3.35M$.
}
\label{fig:THoriz}
\end{figure}

How quickly are our numerical slices flung out of the second copy of
the exterior space? Figure~\ref{fig:THoriz} shows the time a point
takes to reach the inner horizon, parametrized by the point's
isotropic coordinate $r$. The point at $r = M/2$ is at the horizon at
$t = 0$, and so ``reaches'' the inner horizon immediately. The time
for points with $r < M/2$ to reach the inner horizon appears to grow
linearly as we move toward the puncture. Very close to the puncture,
however, the time grows logarithmically, and even the closest grid
point, at $r = M/8192$ and initially at $R = 2049M$, reaches the inner
horizon by about $t = 3.3M$. A similar result is shown in
\cite{Brown:2007tb}. 

\subsection{Late-time behaviour and approach to the stationary solution} 

\begin{figure}[t]
\centering
\includegraphics[width=60mm]{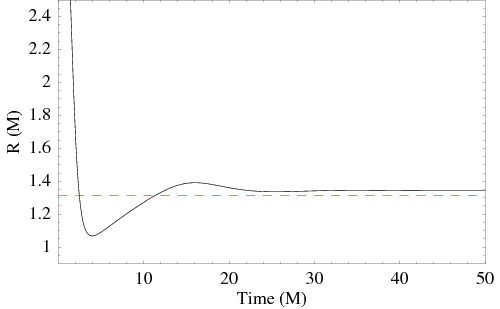}
\caption{The Schwarzschild coordinate $R$ of a point at $r = 3M/32$ as a function of
time. The point overshoots $R_0 = 1.31M$, indicated by a dashed horizontal
line, but for $t > 40M$ settles on a value slightly larger than $R_0$.}
\label{fig:RMotionLate}
\end{figure}

To follow the behaviour of $R(r)$ at later times, we return to our standard 
convergence series simulations. Figure~\ref{fig:RMotionLate} shows $R$ as 
a function of time for a grid point at $r = 3M/32$. We see that the point, having
retreated quickly through the inner horizon, overshoots $R_0 = 1.31M$ (indicated
by a dashed line in the figure), before returning and settling to a value
just larger than $R_0$. We may expect that a point's Schwarzschild coordinate
$R$ can decrease but not increase: a point that falls into the black hole cannot
rise back toward the surface, unless the lapse becomes negative and time
progresses backward. We will return to this surprising behaviour when we 
represent the time development of the numerical slices on a Penrose diagram.

Having described the general behaviour of moving-puncture simulations of
Schwarzschild wormhole puncture data, we would like to verify that the
numerical solution converges to the analytic one. A natural quantity to look
at would be $R$ at the grid points closest to the origin. In Figures~2 and 3
of \cite{Bruegmann:2006at}, we have seen that the metric quantities are
approximately fourth-order convergent for at least $50M$ of evolution. However,
the convergence is not so precise that it is retained in quantities derived
from the evolution variables. In particular, $R = \psi^2 x
\sqrt{\tilde{\gamma}_{zz}}$ does not exhibit clean fourth-order
convergence, and is not suitable for verifying convergence towards the
analytic 1+log geometry. 

As an alternative, we look at the value of $\operatorname{Tr}(K)$ on
the horizon. Although locating the horizon once again requires an
estimation of $R$, the accuracy is much better far from the origin,
and a more systematic analysis of the convergence and accuracy of the
solution is possible.  

\begin{figure}[t]
\centering
\includegraphics[width=40mm]{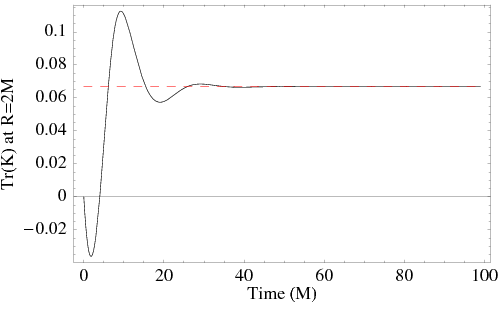}
\quad
\includegraphics[width=40mm]{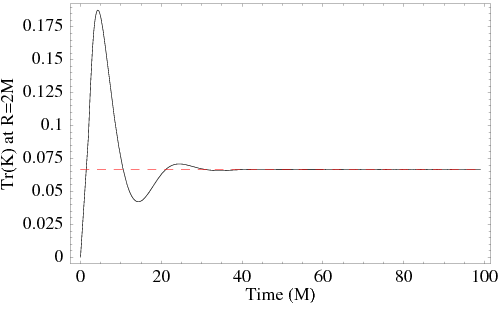}
\caption{Time development of $K$ at the horizon, for simulations with initial lapse
(a) $\alpha = \psi^{-2}$, and (b) $\alpha = 1$.}
\label{fig:KatH}
\end{figure}

\begin{figure}[t]
\centering
\includegraphics[width=40mm]{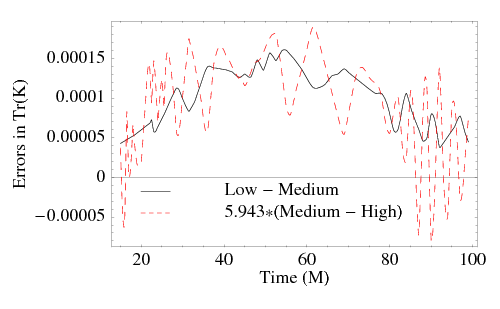}
\quad
\includegraphics[width=40mm]{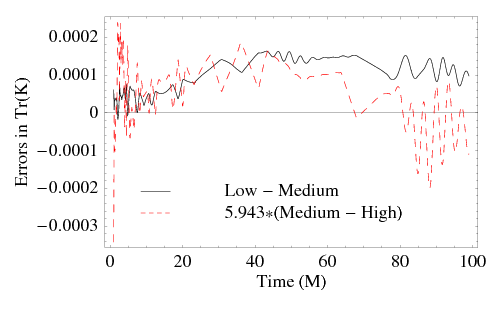}
\caption{Convergence of $K$ at the horizon, for simulations with initial lapse
(a) $\alpha = \psi^{-2}$, and (b) $\alpha = 1$. The
$\alpha = \psi^{-2}$ simulations are not convergent at early times, while the
$\alpha = 1$ simulations lose clean convergence after about $50M$.}
\label{fig:KatHcvg}
\end{figure}

Figure~\ref{fig:KatH} shows the value of $\operatorname{Tr}(K)$ at the horizon for
simulations that begin with $\alpha = 1$ and $\alpha = \psi^{-2}$. In both
cases the value relaxes to the analytic result of $K_H =
0.0668$. Figure~\ref{fig:KatHcvg} shows the convergence of $K_H$ for the
convergence series described earlier. With a ``precollapse'' initial lapse of
$\psi^{-2}$ we see no sign of convergence at early times, but reasonably
clean fourth-order convergence after about $15M$ of evolution. With an initial
lapse of $\alpha = 1$, we see fourth-order convergence after only a few $M$ of
evolution, although the convergence deteriorates at later times.

\begin{figure}[t]
\centering
\includegraphics[width=60mm]{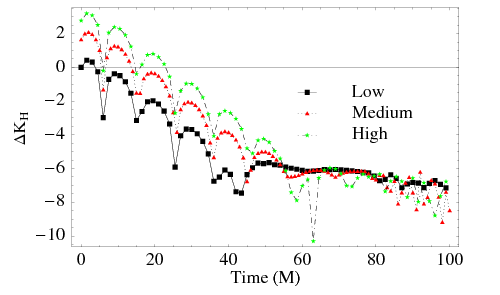}
\caption{Logarithm of the deviation of $K$ at the horizon from the analytic
  value $K = 0.066811$. The deviations are scaled assuming fourth-order
  convergence, and the results indicate that $K$ converges to the analytic
  value with fourth-order accuracy.} 
\label{fig:KatHErrorCvg}
\end{figure}

Figure~\ref{fig:KatHErrorCvg} shows the deviation of $K_H$ from the analytic
value on a logarithmic plot. The values are scaled assuming fourth-order
convergence, and we see clearly that the disagreement between the
numerical $K_H$ and the value for the stationary 1+log slice converge to zero
at fourth-order at late times. This provides strong evidence that the
numerical slice does indeed approach the analytic stationary slice with high
accuracy. 

\subsection{Penrose diagrams of numerical results} 

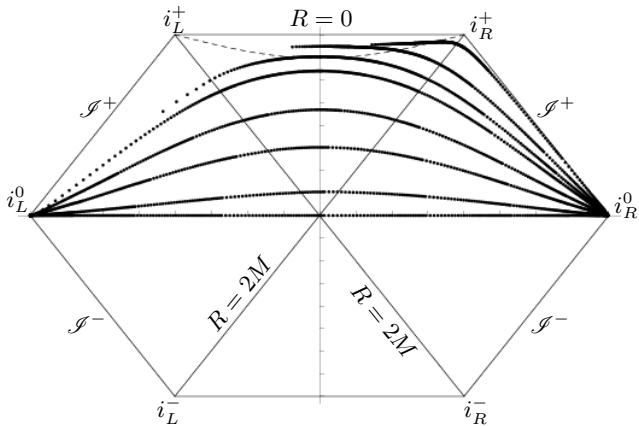
\begin{figure}[t]
\centering
\begin{overpic}[width=80mm]{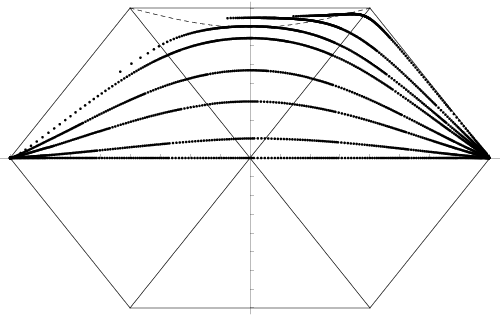}
\put(75,62){$i_R^+$}
\put(24,63){$i_L^+$}
\put(86,48){$\mathscr{I}^+$}
\put(10,48){$\mathscr{I}^+$}
\put(8,12){$\mathscr{I}^-$}
\put(85,12){$\mathscr{I}^-$}
\put(74,-2){$i^-_R$}
\put(23,-2){$i^-_L$}
\put(-2,33){$i_L^0$}
\put(99,32){$i_R^0$}
\put(33,12){\begin{rotate}{51}
  $R = 2M$
\end{rotate}}
\put(55,19){\begin{rotate}{-51}
  $R = 2M$
\end{rotate}}
\put(45,63){$R = 0$}
\end{overpic}
\caption{Penrose diagram produced from numerical data. We can clearly see
the {\it numerical} slice retract from the second asymptotically flat
end. The times shown are $t/M = 0, 0.25, 0.75, 1.25, 2.0, 2.5, 3.5, 8$.}
\label{fig:PenShift}
\end{figure}

We now represent the time development of the numerical slices on Penrose
diagrams, using the technique described in Section~\ref{sec:numpenrose}. 
Figure~\ref{fig:PenShift} shows such a diagram. The initial conditions are
chosen such that the initial data are at $T = 0$ and therefore
correspond to the horizontal line between $i^0_L$ and $i^0_R$. During the
evolution, the slices move upwards symmetrically in the diagram. We
can clearly see that the points move quickly to the right as the slices move up,
and very soon the region near $i^0_L$ (the second asymptotically flat end)
is extremely poorly resolved. In effect the numerical slices lose contact with the
second asymptotically flat end and congregate near the cylinder at $R = 1.31M$,
which is shown by a dashed line.

\begin{figure}[t]
\centering
\begin{overpic}[width=60mm]{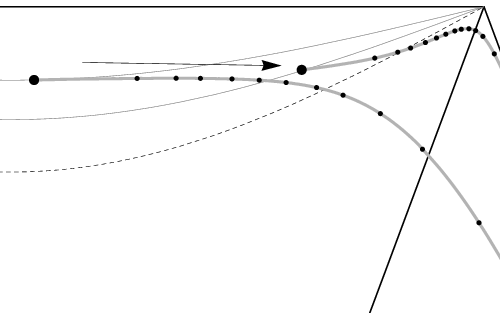}
\put(0,23){$R_0$}
\put(0,34){$R_b$}
\put(-2,48){$R_a$}
\put(39,51){$\beta^r$}
\put(60,62){$R = 0$}
\put(97,62){$i^+_R$}
\put(73,5){\begin{rotate}{67}
$R = 2M$
         \end{rotate}
}
\put(98,20){$t = t_a$}
\put(97,44.5){$t = t_b$}
\end{overpic}
\caption{A close-up of the region near the cylinder at $R_0 = 1.31M$. Although
  the time has elapsed from $t_a \approx 4.5 M$ to $t_b = 13 M$ (with non-negative
  lapse), $R$ at the innermost gridpoint has increased, $R_a < R_b$. } 
\label{fig:PenCloseup}
\end{figure}
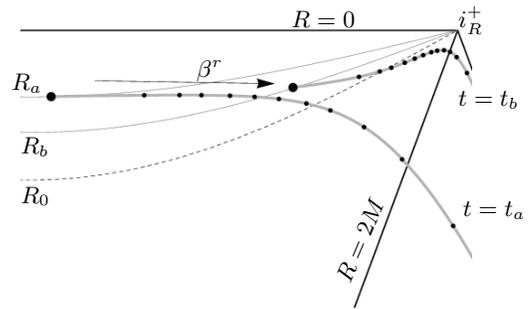

We can also see in Figure~\ref{fig:PenShift} (and as was also clear in 
Figure~\ref{fig:RMotionLate}), that the slices first penetrate $R_0$, before
retreating later to a location just outside $R_0$. We focus on this behaviour
in Figure~\ref{fig:PenCloseup}. 

Our initial reaction to Figure~\ref{fig:RMotionLate} might be that
this is a numerical error: the Schwarzschild $R$ associated with a
point can decrease, but it cannot increase unless the lapse is
negative and time progresses backwards. 
We have already seen that the lapse is everywhere non-negative, so
this behaviour appears to be contradictory. However, the points {\it
  can} move to larger values of $R$ and move forward in time with the
aid of a non-zero shift. Figure~\ref{fig:PenCloseup} illustrates how
this is possible.  

As a further illustration of this point, consider an arbitrary slice through the 
Schwarzschild solution. Choose $\alpha = 0$ and $\beta^R = 1$. The data points
march along the slice, some with decreasing $R$, some with increasing
$R$, depending on the slice we chose. Thus there is no connection
between the change of $R$ and the allowed time vectors.

\begin{figure}[t]
\centering
\begin{overpic}[width=80mm]{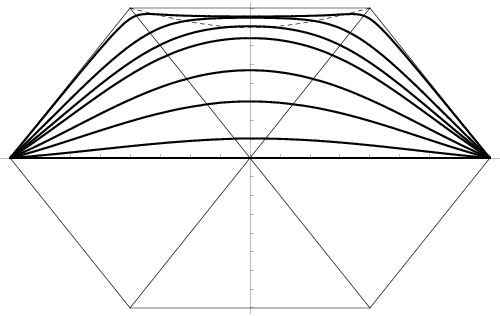}
\put(75,62){$i_R^+$}
\put(24,63){$i_L^+$}
\put(86,48){$\mathscr{I}^+$}
\put(10,48){$\mathscr{I}^+$}
\put(8,12){$\mathscr{I}^-$}
\put(85,12){$\mathscr{I}^-$}
\put(74,-2){$i^-_R$}
\put(23,-2){$i^-_L$}
\put(-2,33){$i_L^0$}
\put(99,32){$i_R^0$}
\put(33,12){\begin{rotate}{51}
  $R = 2M$
\end{rotate}}
\put(55,19){\begin{rotate}{-51}
  $R = 2M$
\end{rotate}}
\put(45,63){$R = 0$}
\end{overpic}
\caption{Penrose diagram of an evolution identical to that used for 
Figure~\ref{fig:PenShift}, except that in this case the shift is zero throughout
the evolution and the data points are joined in the plot. We see that
the numerical slices no longer retract from the second asymptotically
flat end, and now penetrate $R_0$, and stay there. The times shown are
the same as in Figure~\ref{fig:PenShift}. 
}
\label{fig:PenNoShift}
\end{figure}
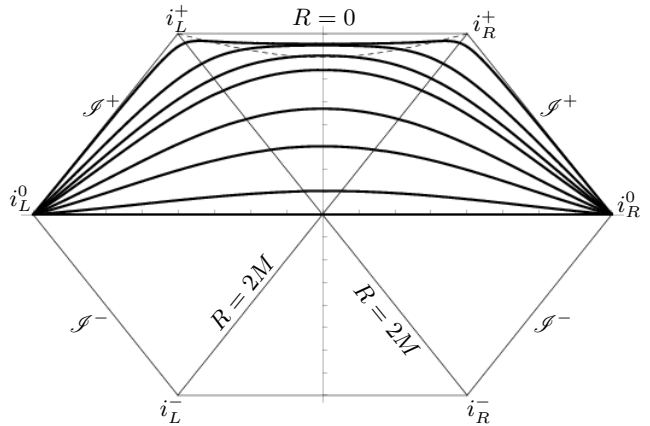

It should also be clear that if we were to run our simulation without a shift, then
the slices would penetrate $R_0$, but would {\it not} be able to retreat to
$R_0$ at later times. This is illustrated in Figure~\ref{fig:PenNoShift}, which was
produced from a numerical simulation with $\beta^i = 0$. This behaviour is in 
direct contrast to what happens in
the case of true maximal slicing (where the maximal slicing condition is imposed
throughout the evolution, and is not approached only asymptotically, as with the 
case of the maximal variant of the 1+log condition, (\ref{eqn:old1plog})), where
the slices approach $R_0 = 3M/2$, but cannot pass through it. Maximal slicing
is an elliptic condition and thus generates ``barriers'', while 1+log slicing is
hyperbolic and so no barriers exist.

Note once again that the slices are isometric across the throat, as shown in 
Figures \ref{fig:PenShift} and \ref{fig:PenNoShift}. The shift only relabels
points within the slices and grid points move accordingly, so the
figures show identical slices covered by different numerical grids.

\begin{figure}[t]
\centering
\begin{overpic}[width=80mm]{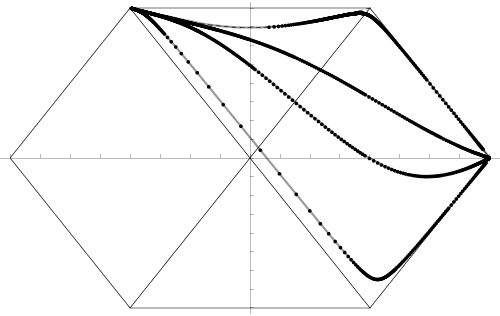}
\put(75,62){$i_R^+$}
\put(24,63){$i_L^+$}
\put(86,48){$\mathscr{I}^+$}
\put(10,48){$\mathscr{I}^+$}
\put(8,12){$\mathscr{I}^-$}
\put(85,12){$\mathscr{I}^-$}
\put(74,-2){$i^-_R$}
\put(23,-2){$i^-_L$}
\put(-2,33){$i_L^0$}
\put(99,32){$i_R^0$}
\put(33,12){\begin{rotate}{51}
  $R = 2M$
\end{rotate}}
\put(55,19){\begin{rotate}{-51}
  $R = 2M$
\end{rotate}}
\put(45,63){$R = 0$}
\end{overpic}
\caption{The numerical data at Schwarzschild times $t = 30, 37.25, 40, 50$M,
with the initial time chosen as $T_0 = -40$M. The analytic solution, evaluated
at the same times, is shown by a solid grey line. We see that the numerical points
lie perfectly on the analytic solution.}
\label{fig:Pen_late}
\end{figure}
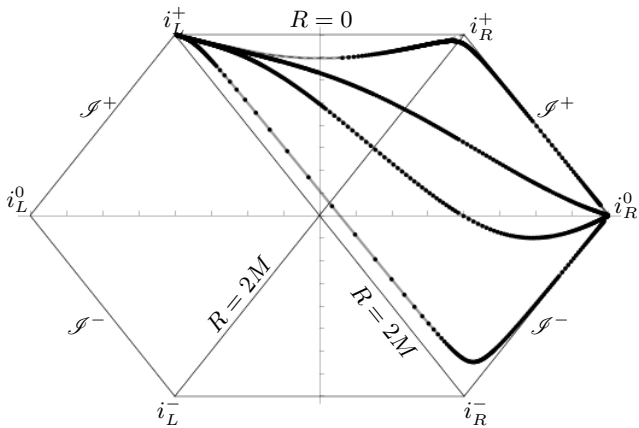

It is difficult to see what happens to the numerical points at late times in 
Figure~\ref{fig:PenShift}, because all of the points bunch up in the upper
right corner of the diagram. We can change this by choosing $T < 0$ for 
the initial slice when constructing the diagram. Figure~\ref{fig:Pen_late} 
shows the slices at numerical times $t = 30, 37.25, 40, 50$M, with the initial
time chosen as Schwarzschild time $T_0 = -40$M. We see clearly that the
slices approach the cylinder at $R = 1.31M$. In addition the figure shows
the analytic solution evaluated at the same times, and we see that the
numerical points lie perfectly on top of the analytic solution, and that since
we have reached the stationary slice, the numerical and Schwarzschild times
coincide.

\section{Numerical evolution of trumpet initial data}
\label{sec:trumpetevolution}

In the previous section we evolved wormhole puncture data with 1+log slicing
and the $\tilde{\Gamma}$-driver shift condition, and found that the numerical
data quickly evolved from a wormhole to a trumpet geometry. In this section we
start with trumpet data. We first show explicitly that they are time-independent
(up to numerical errors, which converge to zero with increasing
resolution). We then demonstrate that, if we alternate between variants of 
1+log slicing during the evolution (in practice (\ref{eqn:old1plog}) and
(\ref{eqn:1plogfull})), the numerical slice  
alternates accordingly between the respective stationary trumpet
geometries. This process also allows us to illustrate how the coordinates can
drift in these evolutions, while invariant quantities remain unchanged.

\subsection{Evolution of time-independent data}

If we evolve the stationary 1+log solution in isotropic coordinates, given in 
Section~\ref{sec:isotrumpet}, we expect the data to be time-independent. 
This is certainly the case when we look at the data by eye: the evolution 
variables do not appear to change at all. 

\begin{figure}[t]
\centering
\includegraphics[width=80mm]{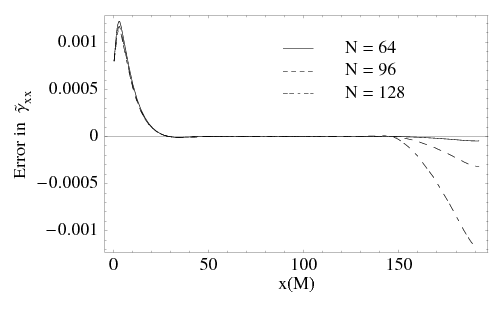}
\caption{The error in $\tilde{\gamma}_{xx}$ (i.e., $\tilde{\gamma}_{xx} - 1$)
  after 50$M$ of evolution of the 
  time-independent trumpet puncture data. The errors for simulations at three
  resolutions are scaled assuming fourth-order convergence, and we indeed see
  that the errors converge to zero at fourth-order.}
\label{fig:gxxError}
\end{figure}

A more systematic test is shown in Figure~\ref{fig:gxxError}, where we show the 
error in $\tilde{\gamma}_{xx}$ at $t = 50M$. The conformal spatial metric is
flat in the stationary data, $\tilde{\gamma}_{ij} = \delta_{ij}$, and so we
can calculate the error by simply evaluating $\tilde{\gamma}_{xx} - 1$. We see
that some error has developed by $t = 50M$, and it has a peak at around $x =
3M$. The errors are scaled consistent with fourth-order convergence, and we
indeed see fourth-order convergence up to around $x = 145M$. Since the outer
boundary is at $x = 192M$, by $t = 50M$ lower-order errors from the outer
boundary will have propagated to around $x = 142M$, and so we do not expect to
see fourth-order convergence beyond this point.

\begin{figure}[t]
\centering
\includegraphics[width=80mm]{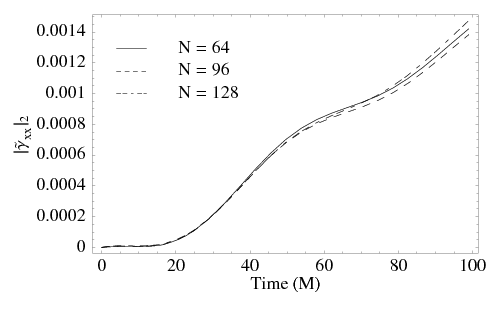}
\caption{The $L_2$ norm of the error in $\tilde{\gamma}_{xx}$ over the course
  of the evolution. The errors again converge to zero at fourth-order.}
\label{fig:gxxNormError}
\end{figure}

In Figure~\ref{fig:gxxNormError} we show the $L_2$ norm of the error in
$\tilde{\gamma}_{xx}$ along the $x$-axis as a function of time. The lines are
once again scaled assuming fourth-order convergence to zero. We see reasonably
clean fourth-order convergence, which appears to deteriorate slightly near the
end of the simulation, although by this time lower-order errors from the outer
boundary will have contaminated the solution, as is clear in
Figure~\ref{fig:gxxError}.   

These figures indicate that the data are indeed time independent, up to small
numerical errors that converge to zero at the expected rate. Since the
analytic value of $\tilde{\gamma}_{xx}$ is unity, 
we can easily calculate the percentage error from the figures: we see that at
$t = 50M$ the largest error in $\tilde{\gamma}_{ij}$ is 0.12\%, for the
lowest-resolution simulation. As such, we see that these
data provide an excellent testing ground for the accuracy of a numerical
code. This point needs to be emphasized. The moving-puncture approach as used
here is currently the most popular method for simulating black-hole
binaries. For such a code there is {\it no} analytic black-hole solution that
can be used to test the code, except for the 1+log and stationary maximal
trumpet solutions presented here and in
\cite{Hannam:2006xw,Baumgarte:2007ht}. These analytic solutions could be
invaluable not only for testing a new code, but also in analysing and reducing
the sources of error in current codes.

\subsection{Alternating slices}

We start with 1+log trumpet data, as in the previous section. We evolve the
data for $t = 22.5M$, and then switch the slicing evolution equation from
(\ref{eqn:1plogfull}) to (\ref{eqn:old1plog}), i.e., we change to the slicing
condition that asymptotes to maximal slicing. After a further $50M$ of
evolution, at $t = 72.5M$, we switch back to the original slicing condition,
which, assuming robustness of the method, should asymptote back to the
stationary 1+log solution. 

\begin{figure}[t]
\centering
\includegraphics[width=80mm]{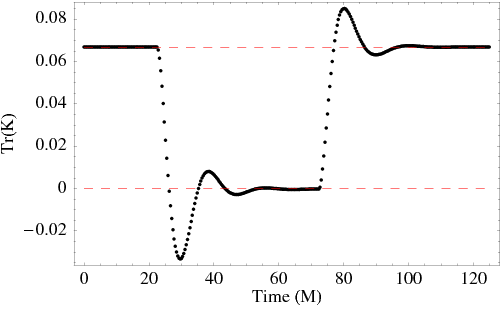}
\caption{The value of $K$ at the horizon, for a simulation where the slicing
  condition alternates between standard 1+log and asymptotically maximal
  1+log, Eqns.\ (\ref{eqn:1plog}) and (\ref{eqn:old1plog}). The dashed
  horizontal lines indicate the respective analytic values of $K$ at the
  horizon.}
\label{fig:alternatingK}
\end{figure}

Figure~\ref{fig:alternatingK} shows the value of $K$ at the horizon for this
simulation. The value is $K = 0.0668$ on the horizon in the initial data, and
remains at this value for the first $22.5M$ of evolution. Then, when the
slicing condition changes, $K$ quickly evolves towards $K = 0$. Within about
$30M$ we have $K \approx 0$. At $t = 72.5M$, the slicing condition is changed
again, and within another $30M$ the slice has settled back to the stationary
1+log value. Simulations were performed with the same low, medium and high
resolutions as the puncture data case in Section~\ref{sec:numresults}. At $t =
125M$, when the simulations ended, the respective values of $K$ on the horizon
were $0.0679, 0.0670, 0.0669$. 

These results illustrate the robustness of the moving-puncture method to locate
the appropriate stationary 1+log slice. Potentially more challenging tests have
also been performed using excision initial data with the interior
filled in, and the method is again seen to be robust
\cite{Brown:2007pg,Etienne:2007hr,Faber:2007dv}. 

\begin{figure}[t]
\centering
\includegraphics[width=80mm]{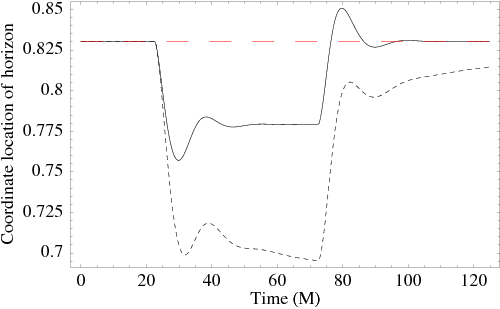}
\caption{The coordinate location the horizon, for two simulations with an
  alternating slicing condition. The dashed line is for a simulation with the
  standard choice of $\eta = 2/M$. The solid line is for a simulation with
  $\eta = 0$. In this case, the horizon location almost returns to its
  original value when the solution returns to the 1+log stationary geometry.}
\label{fig:alternatingH}
\end{figure}

In Figure~\ref{fig:alternatingH} we show the coordinate distance $r$ of the
horizon from the origin. For the stationary 1+log data the horizon is at $r = 
0.8304M$, and deviates by no more than $0.0014$~\% in the first $22M$ of
evolution in the highest-resolution simulation. When the slicing condition is
switched to asymptotically maximal 1+log, and then back to standard 1+log, we
see that the horizon does not return to its original position, at least not on
the same time scale as the geometry returns to the stationary 1+log
geometry. (See the dashed line in the figure.) This illustrates that, although
the numerical slices quickly approach a stationary {\it geometry}, the
coordinates may still drift. This effect is at least partially due to the
damping parameter $\eta$ in the $\tilde{\Gamma}$-driver shift condition. If we
repeat the simulation with $\eta = 0$, we produce the solid line in
Figure~\ref{fig:alternatingH}: now the horizon location returns to its
original location to a comparable accuracy that coordinate-invariant
quantities return to the stationary 1+log solution. 

\section{Discussion}

\subsection{Setting wormholes and trumpets in motion}

One of the directions for future work suggested by our results (and
already proposed in Paper I) is the construction of trumpet (as
opposed to wormhole) initial data for black-hole binaries. 

Trumpet black-hole binary initial data would have the advantage that 
our numerical code would not need to evolve the fast transition from 
wormholes to trumpets. In addition, current binary puncture initial
data start with zero speed across the numerical grid. When the
simulation begins, the 1+log/$\tilde{\Gamma}$-driver gauge conditions
both transform the wormholes into trumpets, but also generate an
advection component to the shift vector, which moves the trumpets
across the grid (Paper I). One could attempt to choose a ``better''
initial shift, so that the wormholes move from the outset, but if we
begin with wormhole puncture data, there is no way to prevent this
shift changing nontrivially as the wormholes evolve to
trumpets. Although one could produce a ``best'' initial shift by some
trial-and-error process, a more attractive proposal would be to start
with trumpet data, and hope to find a procedure to choose an initial 
shift that imbues the boosted trumpets with their appropriate
coordinate speeds. We would then hope that, {\it from the outset},
almost all of the evolution of the data would represent the evolution
of {\it physical} quantities (the black holes' motions and the
evolution of their spins, surface geometry, etc.) and mere gauge
evolution would be minimised. This might help reduce the gauge
component to the black-hole motion seen in, for example, the first few
hundred $M$ of evolution in Figure~18 in
\cite{Hannam:2007ik}. Further, and more practically, we would hope
that such data would reduce the initial noise in wave quantities (for
example, Figures 1, 2, 4 and 6 in \cite{Hannam:2007ik}).  

Before we become too optimistic, however, we should emphasize that the
bulk of the initial noise in most numerical simulations probably comes
from the burst of junk radiation present in the initial data. For
example, the excision data evolved in \cite{Boyle:2007ft} already have
most of the properties we have just advertised: the gauge is not
expected to evolve significantly in the early stages of the
simulation, and the initial shift is precisely that which should cause
the black holes to follow a quasi-circular inspiral. As we would hope,
the black-hole motion is indeed smooth from the outset. Nonetheless,
noise still reduces the accuracy of some wave quantities at early
times (see, for example, Figure~7 of \cite{Boyle:2007ft}), and the
most likely culprit is the junk radiation, which has a similar
magnitude to that in moving-puncture simulations. We expect that the
ideal initial data for moving-puncture simulations would produce
minimal junk radiation {\it and} be in trumpet form.

\subsection{Summary}

We have extended the analysis of the behaviour of the analytical and
numerical slices in moving-puncture simulations of the Schwarzschild
spacetime that we began in Paper I. For a general form of the 1+log
slicing condition (\ref{eqn:1plog}) we have derived the stationary
Schwarzschild trumpet solution: the slice extends from spatial
infinity to an infinitely long cylinder, or trumpet, with a throat at
some finite radius $R_0$. When the parameter $n$ in this condition is
set to $n = 2$, we obtain the solution given in Paper I with $R_0 =
1.31M$. In the limit $n \rightarrow \infty$ we recover the maximal
trumpet solution \cite{Estabrook73,Reinhardt73}, with $R_0 = 1.5M$.

For a given choice of the 1+log slicing condition, there is a
unique regular stationary solution. In numerical simulations that
apply the moving-puncture technique to wormhole puncture Schwarzschild
initial data, {\it and} use the $\tilde{\Gamma}$-driver shift
condition, the numerical slice quickly evolves to the stationary
slice. This cannot happen to the analytic slice: this must remain
connected to the two asymptotically flat ends in the wormhole data. It
also cannot happen with numerical data if the shift is zero. However,
the $\tilde{\Gamma}$-driver shift condition generates a shift that
stretches the numerical slice such that all of the numerical points
extremely quickly move onto the stationary 1+log slice; the
non-stationary part of the slice no longer contains any grid
points. The stretching of the slice is so extreme that no matter how
many numerical points we place near the puncture (so long as there is
no point {\it on} the puncture) that point will quickly move onto the
stationary slice. Even a grid point initially at $r = M /
8192$ and $R \approx 2000M$ on the second copy of the exterior space,
passes through $R = 2M$ in less than $3.5M$ of evolution, and soon
after settles near $R = 1.31M$.

An alternative to wormhole puncture data are {\it trumpet} puncture data. We
have transformed the stationary 1+log solution to isotropic coordinates, and
shown that these data are indeed time-independent when evolved, up to small
numerical errors. In addition, we have shown that the numerical data can
easily change from one stationary geometry to another, if the 1+log condition
is changed during the evolution, indicating a certain robustness of
the method. Finally, we are able to see clearly that
although the data approach a stationary slice, the numerical coordinates may
still drift. 

The realization that the moving puncture method is really based on trumpet
puncture data, and that this type of geometry naturally avoids most of the
unphysical regions of spacetime in 
black hole evolutions, establishes a new paradigm for the numerical evolution
of black holes and suggests many directions of possible future research. One
of the most promising is the construction of trumpet initial data for
black-hole binaries, as discussed above. Additionally, one may use the
stationary data to make precise tests of numerical codes, 
for example to improve treatment of mesh-refinement boundaries and
outer boundaries, to determine the resolutions necessary to most
accurately resolve black-hole spacetimes, and to explore different
gauge choices. The stationary solution also provides an ideal
background for mathematical studies of the stability properties of the
BSSN/moving-puncture system and other evolution systems, that could
even be particularly tailored to evolve trumpet puncture data.

\acknowledgments

This work was supported in part by DFG grant SFB/Transregio~7
``Gravitational Wave Astronomy''.  In addition, M. Hannam and
N. \'O~Murchadha were supported by SFI grant 07/RFP/PHYF148.
S. Husa is a VESF fellow of the European Gravitational Observatory
(EGO).
We thank
the DEISA Consortium (co-funded by the EU, FP6 project 508830), for
support within the DEISA Extreme Computing Initiative
(www.deisa.org). Computations were performed at LRZ Munich and the
Doppler and Kepler clusters at the Institute of Theoretical Physics of
the University of Jena.

\bibliography{refs}

\end{document}